\documentclass[12pt]{aastex}

\newcommand{\Msun}{\mbox{ M}_{\odot}}
\newcommand{\gae}{\mathrel{>\kern-1.0em\lower0.9ex\hbox{$\sim$}}}
\newcommand{\lae}{\mathrel{<\kern-1.0em\lower0.9ex\hbox{$\sim$}}}
\newcommand{\kms}{km~s$^{-1}$}

\newcommand{\CIV}{\ion{C}{4}\,$\lambda\lambda\,1548.2,\,1550.8$\ } 
\newcommand{\SiIV}{\ion{Si}{4}\,$\lambda\lambda\,1393.7,1402.8$\ }

\lefthead{Boroson et al.}

\begin{document}

\title{Hot Outflowing Gas from the X-ray Binary Hercules X-1}

\author{Bram Boroson}
\affil{College of Wooster, Wooster, OH 44691; bboroson@acs.wooster.edu}

\and

\author{Timothy Kallman}
\affil{Goddard Space Flight Center, Greenbelt, MD 20771;
tim@xstar.gsfc.nasa.gov}

\and

\author{Saeqa Dil Vrtilek}
\affil{Harvard-Smithsonian Center for
Astrophysics; svrtilek@cfa.harvard.edu}


\begin{abstract}

We present a unified picture of outflowing gas from the X-ray binary
system Hercules X-1/HZ Herculis.  We suggest that the outflowing gas 
(a wind) causes UV
emission seen in mid-eclipse, narrow UV absorption lines, and broad UV
P~Cygni lines.
Observations with the FOS and
STIS spectrographs on the Hubble Space Telescope (HST) show UV emission
lines in the middle of X-ray eclipse, when the X-ray heated atmosphere of
the normal star and accretion disk should be entirely hidden from view.  
Narrow absorption lines (FWHM$\approx50$~km~s${-1}$) blueshifted by
500~km~s$^{-1}$ during observations in 1998 and by 400~km~s$^{-1}$ during
observations in 1999 were seen from $\phi=0.-0.3$.  The line velocity
was constant to within 20~km~s$^{-1}$.  The P~Cygni 
profiles
from Hercules~X-1 have optical depths $\tau\lae1$ with a maximum
expansion velocity of $\approx600$~km~s$^{-1}$, and are seen in the
resonance lines \ion{N}{5}$\lambda\lambda1238.8,1242.8$,
\SiIV, and \CIV.
We discuss whether this wind originates in the accretion
disk or on the companion star, and how the relevant ions
can survive X-ray ionization by the neutron
star.

\end{abstract}

\section{Introduction}

Hercules~X-1/HZ~Herculis is an X-ray binary consisting of a 1.24 second
pulsar in an eclipsing 1.7 day orbit with a $\sim2\Msun$ mass normal
companion. As a result of its many periodicities, it is one of the most
frequently observed X-ray binaries.  The X-rays vary over a 35~day cycle;
an $\approx8$ day ``Main-on'' state and $\approx4$ day ``Short-on'' state
(in which the observed X-ray flux is reduced by a factor of $\approx3$)
are separated by half of a 35-day phase.  Outside of these states the
X-ray flux is a few percent of that seen in the Main-on state. The X-ray
modulation is not due to a change in the total X-ray output, as during
the X-ray low period, the
optical magnitude continues to vary over the 1.7~day orbit because of 
X-ray
heating of the companion star.  Instead, the 35-day variation probably
results from obscuration of the central source by an accretion disk which
wobbles over a 35-day period due to an unknown cause.  X-ray absorption
dips occur at a period of 1.65~days, near to, but significantly greater
than, the 1.62~day beat period between the 1.7 and 35 day periods (Crosa
\&\ Boynton 1981; Scott \&\ Leahy 1999). 

The far UV wavelength range is ideal for investigating the
accretion disk or winds in the system, as it includes many strong
resonance lines (including
\ion{N}{5}$\lambda\lambda1238.8,1242.8$,
\ion{Si}{4}$\lambda\lambda1393.8,1402.8$, and
\ion{C}{4}$\lambda\lambda1548.2,1550.8$), and the continuum emission from
the disk peaks in the far UV.  Her~X-1 was observed extensively with IUE
(Dupree et al. 1978; Gursky et al. 1980; Howarth \&\ Wilson 1983a,b; Boyle
et al. 1986; Vrtilek \&\ Cheng 1996).  These investigations
emphasized the orbital and 35-day variation in the line and continuum
flux.

With a greater collecting area, HST could observe more subtle effects in
the UV spectrum.
Anderson et al. (1996) discovered that,
surprisingly, even in the middle of eclipse ($\phi=0.995-0.006$), emission
lines of \ion{N}{5}$\lambda1240$, \ion{Si}{4}$+$\ion{O}{4}]$\lambda1400$,
\ion{N}{4}$\lambda1487$, and \ion{C}{4}$\lambda1550$ persist at
$\approx1$\%\ of the flux seen at orbital phase $\phi=0.5$, inferior
conjunction of the neutron star.
Photoionization models implied that
the gas visible in mid-eclipse has a density $n_e \lae 10^{11}$~cm$^{-3}$,
a temperature 15,000~K$<T<$33,000~K, an ionization parameter $0\lae\log
\xi\lae1$ and an absorbing column density of $N_p \gae 10^{19}$~cm$^{-2}$.  
The absence of detectable \ion{O}{5}$\lambda1371$ emission and the
presence of \ion{N}{4}$\lambda1487$ alone clearly rule out
$n_e\gg 10^{11}$~cm$^{-3}$.

Evidence at other wavelengths supports the existence of this gas.  
Extended hot gas has been invoked to explain X-ray spectra during the low
state (Mihara et al. 1991), and at mid-eclipse (Parmar et al. 1985;
Mavromatakis 1993; Choi et al. 1994).  

UV observations of Her~X-1 with the Goddard High Resolution Spectrograph
(GHRS) have shown that during one HST orbit at Her~X-1 orbital phase
$\phi=0.80$, the \ion{N}{5} doublet component at 1242.8\AA\ was stronger
than the component at 1238.8\AA, which should be impossible, as the
oscillator strength of the 1238.8\AA\ line is twice as strong (Boroson et
al. 1996).  One possible explanation for this is that there was
blue-shifted P~Cygni absorption associated with each doublet component,
and that the 1242.8\AA\ line could absorb the 1238.8\AA\ line, but not
vice-versa.  However, this was only an indirect argument for a wind, as
the blue-shifted absorption was never directly seen, and it is difficult
to account for the presence of ions such as \ion{N}{5} in a wind that may
be ionized by Her~X-1, with an X-ray luminosity
$L_x\approx3\times10^{37}$~erg~s$^{-1}$.

The UV line profiles seen with the HST~GHRS and Space Telescope Imaging
Spectrograph (STIS) do not have the double-peaked shape expected from a
Keplerian accretion disk (e.g. Smak 1969).  The line profiles have been
successfully fit with a model in which an accretion disk wind is
responsible for the single-peak line shape (Chiang 2000, submitted).

Accretion disk winds in Cataclysmic Variables are commonly observed
through P~Cygni lines (Drew 1997).  Such winds
typically reach velocities of 3000-6000~km~s$^{-1}$ with mass-loss rates
of $\dot{M}\lae0.05\dot{M_{\rm acc}}$ where $\dot{M_{\rm acc}}$ is the
white dwarf mass accretion rate.  Only recently has there been any 
evidence
for winds in low-mass X-ray binaries, however.  Most of this evidence is
from IR spectroscopy; for example Chakrabarty, van Kerkwijk, \&\ Larkin 
(1998) have shown that there is
an outflow of 250~km~s$^{-1}$ with $\dot{M}\sim10^{-6}\Msun$~yr$^{-1}$ in
GX~1+4/V2116
Ophiuchi.  IR P~Cygni lines have also been reported from Sco~X-1 and
GX~13+1 (Bandyopadhyay et al. 1999).  Recently, {\it Chandra} has revealed
P~Cygni profiles in X-ray lines in the unusual X-ray binary Circinus~X-1
(Brandt \&\ Schulz 2000).

There have been some theoretical investigations of winds in low-mass X-ray
binaries, with special attention given to Hercules~X-1 (Arons 1973;
Davidson \&\ Ostriker 1973; McCray \&\
Hatchett 1975; Basko et al. 1977; London, McCray, \&\ Auer 1981).  No
consensus has been reached from the various numerical and analytic models
of Hercules~X-1 on
whether a wind that forms from the companion's atmosphere can provide all
of the mass needed to power the X-rays from the neutron star (the
``self-excited wind'' model).  X-ray heated and self-excited winds can
have important implications for the late evolution of low-mass X-ray
binaries (Ruderman et al. 1989; Iben, Tutokov, \&\ Fedorova 1997).

Here we consider the FOS eclipse spectrum along with the results of more
recent observations
of the emission lines seen in eclipse and of narrow
UV absorption lines (Boroson et al. 2000a, Vrtilek et al. 2001).  We
present the first attempt to model the P~Cygni profiles
of \ion{C}{4}, \ion{Si}{4}, and \ion{N}{5}.
We propose a common origin for the emission lines seen in eclipse, the
narrow absorption, and the P~Cygni absorption.
We calculate physical parameters of the gas, speculate
on its origin as a wind from the star or disk, and make predictions for
further observations.

\section{Observations}

In 1998 and 1999, we carried out multiwavelength studies of Hercules~X-1
using
the HST~Space Telescope Imaging Spectrograph (STIS), the Rossi X-ray
Timing Explorer (RXTE), the Extreme Ultraviolet Explorer (EUVE), and
ground-based observatories, including the Keck Telescope. Empirical models
for the UV emission lines observed during the first segment of our
campaign,
in July of 1998, have been reported in Boroson et al. (2000a). 
The second segment of the campaign took place in July of 1999 during
an ``anomalous low'' state in which the
X-ray flux was two orders of magnitude lower than in the expected
``main-on'' state, although accretion continued to take place (Parmar et
al. 1999).

The STIS instrument design is described by Woodgate et al. (1998), and the
in-orbit performance of the STIS is described by Kimble et al. (1998).
Due to the sensitivity of STIS to the radiation exposure in the South
Atlantic Anomaly, STIS observations are possible only for those HST orbits
that do not pass through the SAA.  As a result, all of our STIS
observations follow a pattern of 5 HST orbits of observations, followed by
10 HST orbits without observations.

A log of all of the HST far UV (1200--1700\AA) observations of
Hercules~X-1 is shown in Table~1 (in the rest of the paper we will refer
to observations by the labels given in Table~1).  The HST observations of
Her~X-1 to date
include 48~files for a total of 102~ksec. (Some files contain
less than a complete HST orbit, as the SAA interrupted some orbits and 
in some cases detectors were switched in mid-orbit.)
Observations extending into
mid-eclipse have all been at low resolution, employing either the
FOS~G130H
(resolution $R=\lambda/2\Delta\lambda\approx 400$) or the STIS in
low-resolution mode, (G140L, $R=960-1440$).  The 1994
campaign described in Boroson et al. (1996) used the GHRS, alternately
in low-resolution (G140L, $R\approx1000$) and medium-resolution
(G140M, $R\approx20,000$) modes.  The GHRS observations used RAPID readout
mode for a time resolution of 0.1 seconds (Z2HP0202) or 0.35 seconds (all other exposures).

The remaining 29 HST orbits employed the STIS E140M grating for high
resolution
echelle spectroscopy.  This provides a resolving power of $R=45,800$
(6~\kms). For these exposures, we used the TIME-TAG mode on the STIS,
which stamps each photon detected with a time accurate to $125$~$\mu$sec.
The UV continuum observed during the main-on state with the GHRS
was found to pulsate with an rms amplitude of 0.4\%\ at $\phi=0.56$ and
0.7\%\ at $\phi=0.83$ (Boroson et al. 1996).  During the anomalous low
period, reprocessed UV pulsations were not seen, although UV quasiperiodic
oscillations (QPOs) at 8 and 45 mHz were discovered (Boroson et al.
2000b).

\subsection{Mid-eclipse lines}

The 1998 STIS observations included low-resolution exposures at
$\phi=0.985$ and $\phi=0.986$ (sub-exposures of observation root name
O4V401050.) Figure~1 compares these spectra with mid-eclipse spectra
observed with the FOS.  Both observations detect the same lines, except
for the \ion{He}{2}$\lambda1640$ line, which is out of the wavelength
coverage of the FOS. The detection of this line in mid-eclipse is not
surprising, as Anderson et al. showed that the optical
\ion{He}{2}$\lambda4686$ line also persists through mid-eclipse.

This STIS observation at mid-eclipse was at long-term phase $\Psi=0.64$,
whereas the FOS observation was at $\Psi=0.76$.  We use the definitions of
the long-term phase given by Scott \&\ Leahy (1999) from their analysis of
RXTE All Sky Monitor data.  The phase origin of the long-term variability,
$\Psi=0.00$, corresponds to the time of X-ray turn-on, while the length of
the long-term period is assumed to be 34.853 days, or exactly 20.5 times the 
orbital period.
The ``Main High'' part of the cycle then lasts from $\Psi=0$ through
$\Psi=0.31$, following by a low state through $\Psi=0.57$, and a ``Short
High'' state from $\Psi=0.57$ through $\Psi=0.79$.  The X-rays are then in
the low state until the start of the next cycle (that is, they are low from
$\Psi=0.79$ through $\Psi=1.00$.

\subsection{Absorption lines}

We list all detections of variable stellar absorption lines (P~Cygni and
narrow absorption) in Table~2.  For a feature to be considered a narrow
absorption line, we require that it be contained entirely within a full
velocity width of 150~km~s$^{-1}$.  There is transient evidence for other
narrow absorption lines, but we limit our attention to lines at blueshifts
near $400-500$~km~s$^{-1}$.  The significance of a feature is higher when we
have an expectation of its wavelength.  When both narrow and P~Cygni
absorption is present, as in the 1238.8\AA\ line in observation O4V408040,
we consider the P~Cygni line to consist of the total absorption.

We integrated the absorbed flux to
determine the equivalent width W$_\lambda$, velocity range, and centroid
velocity.  Repeating this procedure 5 times for different choices of
continuum level and slope allowed us to estimate errors on these
parameters.

Narrow absorption lines were detected in both \ion{N}{5} and
\ion{C}{4}, as shown in Figure~2 and Figure~3.  We have averaged the
spectra over 5~contiguous HST orbits.  At $\phi=0.14,0.27$ the narrow
absorption lines are prominent, while at $\phi=0.59,0.70$ their presence
is obscured by broader absorption, reminiscent of P~Cygni lines, which
we discuss in \S\ref{sec:pcygni}.

It is important to note, first, that these absorption lines are stable in
velocity over orbital phase.  We compare the \ion{N}{5} absorption lines
at $\phi=0.102$ (O4V406020) and at either
$\phi=0.210$ or $\phi=0.288$ (O4V408010 or
O4V408030) and find that the velocities agree to within
$\pm20$~km~s$^{-1}$.  Between these phases, we would expect the velocity
of the neutron star to differ by 60~km~s$^{-1}$.

Second, we find that the lines are stable over periods of days. The
absorption lines at O4V406020 and O4V408040, 2 days apart, agree also to
within $\pm20$~km~s$^{-1}$.

We have already reported the detection of absorption lines in the 1998
STIS campaign, in observations with root names O4V4030x0 (Boroson et al.  
2000a). We find that the absorption line
velocities are {\it not} stable over periods of a year.  The lines
appeared at a (heliocentric) velocity of $-500$~km~s$^{-1}$ in 1998
(O4V4030x0) and at a (heliocentric) velocity of $-400$~km~s$^{-1}$ in 1999
(O4V4060x0 and O4V4080x0).  

The absorption line equivalent widths are variable as well.  As we showed
in Boroson et al. (2000a), the differences between succesive spectra
do not show an increase in the amount of absorbed flux, as they would if
the equivalent width were constant.  For observations with root name
O4V4060x0, the equivalent widths of the 1238.8\AA\ and 1242.8\AA\ lines
were 0.30\AA\ and 0.18\AA.  For O4V4080x0, these values were 0.18\AA\ 
and 0.12\AA.  For the 1998 observations (O4V4030x0), we found
equivalent widths $W_\lambda=0.12, 0.08$\AA.

It is difficult to determine whether the narrow absorption lines are
present for $\phi=0.5-0.6$, as they would overlap with broad P~Cygni
absorption
(\S\ref{sec:pcygni}).  The narrow absorption lines do {\it not} appear in
spectra at $\phi=0.7-0.8$.

\subsection{P Cygni profiles\label{sec:pcygni}}

The emission lines of \ion{C}{4} show very clearly (Figure~4a)
blue-shifted absorption that peaks at a velocity $v\approx300$~km~s$^{-1}$
and extends to $v\approx600$~km~s$^{-1}$.  The absorption reaches a
maximum optical depth $\tau=0.8$ (here, by ``optical depth'' we mean simply 
$-\log_e(F/F_c)$.)
The \ion{Si}{4} and \ion{N}{5} lines also show evidence for P~Cygni
absorption (Figure~4b,c), although the absorption in these cases is not
as clear as for \ion{C}{4}.

These P~Cygni signatures were seen only for $\phi=0.50\pm0.25$ 
in the 1999 data.  The two HST orbits with $\phi=0.650,0.685$
in the 1998 data showed no P~Cygni signatures.

Although the absorption feature appears directly to the blue of the narrow
emission feature, and this is the classic form for a P~Cygni line from an
expanding envelope, the two features are not necessarily related.  We show
in the next section that we observe a similar broad absorption feature
when the narrow emission is not seen. 

\subsection{P Cygni absorption without emission at $\phi=0.10$}

At $\phi=0.10$ (O4V406020) we see an apparent gap in the
broad emission lines of \ion{N}{5} from velocities $v=-250$~km~s$^{-1}$ to
$v=0$
(heliocentric).  This gap is filled in during the next HST orbit
(at $\phi=0.14$), as seen in Figure~5a.

The evidence suggests that this is an absorption feature, and not a 
deficit in emission at certain wavelengths 
(due for example, to a double-peaked line shape typical of
emission from accretion disks).  We show in Figure~5b that the
\ion{O}{5}$\lambda1371$
line, which for reasons of atomic physics is highly unlikely to cause
absorption, does not have a gap in its emission from $v=-250$~km~s$^{-1}$
to $v=0$.  Similarly, Figure~5e shows no such clear gap in the
\ion{He}{2}$\lambda1640$ line, which connects excited levels.  There is
evidence for absorption in \ion{C}{4} (Figure~5d) and \ion{Si}{4}
(Figure~5c).  For these lines, the absorption is blended with the strong,
constant interstellar absorption near $v=0$.

The July 1998 STIS observations showed no such gap in
emission (or absorption) near $\phi=0.10$.

That the absorption is blueshifted at $\phi=0.5$ suggests that the gas
stream between the two stars might be responsible.  Doppler tomograms,
however, are not consistent with this interpretation (H. Quaintrell,
personal communication).  Furthermore, we have described the absorbing gas
as uniformly covering the emitting region with an optical depth
$\tau\approx 1$ from $\phi=0.3$ through $\phi=0.7$.  To find the minimum
portion of the emission region that needs to be covered, allowing the
covering to be non-uniform, we assume that the absorption is saturated.  
Thus we find that at least $\approx60$\%\ of the emission region needs to
be covered in order to reproduce the observation.  It does not seem likely
that portions of the gas stream at the same projected velocity could cover
at least 60\%\ of the heated face of HZ~Her throughout the phase interval
$\phi=0.3$ to $\phi=0.7$.

\section{Identification of Line Regions: Emission, Absorption, and
P~Cygni Lines}

First, we summarize what we know about the emission, absorption, and
P~Cygni features, and their orbital and random
variability.

\begin{itemize}
\item{Emission}: Narrow (FWHM$\approx150$~km~s$^{-1}$) emission lines
probably caused by the heated atmosphere of HZ~Her are seen from
$\phi=0.2$ to $\phi=0.8$, and peak at $\phi=0.5$.  Broad
(FWHM$\approx800$~km~s$^{-1}$) emission
lines probably caused by the accretion disk are seen at a constant flux
from $\phi=0.15$ to $\phi=0.85$, and gradually diminish from $\phi=0.9$
into mid-eclipse (Boroson et al. 2000a).  Emission lines are seen in
mid-eclipse with $\approx1$\%\ of the $\phi=0.5$ flux, but their
velocities and widths are not known.
\item{Absorption}: Observed from $\phi=0.1$ to $\phi=0.3$ in both 1998 and
1999.  Blue-shifted by $500$~km~s$^{-1}$ in 1998 and $400$~km~s$^{-1}$
(heliocentric) in 1999.
\item{P~Cygni}: Observed with maximum velocity $v_{\rm
max}=600$~km~s$^{-1}$ from $\phi=0.25$ to $\phi=0.70$ in 1999.
Possibly observed with $v_{\rm max}\gae1000$~km~s$^{-1}$ at $\phi=0.80$ in
1994.  Possibly observed with $v_{\rm max}=250$~km~s$^{-1}$ at $\phi=0.10$
in 1999.
\end{itemize}

We first try to place the origin of the P~Cygni lines within the
system, then quantify the mass-loss rate and discuss the cause of the line 
variability.
Finally, we relate the emission lines seen in eclipse to the narrow and
P~Cygni absorption lines.

We can set some limits on the origin of the P~Cygni absorption. As
the accretion disk contributes $\lae10$\%\ of the continuum near
$\phi=0.5$ during the anomalous low state in 1999, the absorption, which
is deeper than 10\%, must cover the normal star.  At $\phi=0.10$, the
heated face of the normal star is not yet visible, and the absorption
clearly covers the broad emission line which we have attributed to the
accretion disk (Boroson et al. 2000a).  If we were seeing the results of a
disk wind covering the face of HZ~Her at $\phi=0.5$, we would expect to
see both blue and red-shifted absorption, as the wind would expand away
from the neutron star.  Thus it seems unlikely that we are observing the
effects either of a wind close to the surface of HZ~Her or a wind from the
accretion disk.  This leaves as a most likely explanation that the P~Cygni
absorption is due to a wind from the heated face of HZ~Her that has
expanded beyond the radius of the accretion disk.

We quantify the mass-loss rate of such a wind that is required to
reproduce the observed P~Cygni line optical depth.  As the
\ion{C}{4}$\lambda1548.2$ line showed the clearest P~Cygni profile, we
concentrate on that line.  According to the Sobolev approximation,
\begin{equation} \label{eq:sobolev} \tau=\frac{\pi e^2}{m_e c} n_p a_C
g_{\rm C\,IV} f \lambda (dv/dr)^{-1}, \end{equation} 
where $\lambda$ is
the rest wavelength, $a_C$ is the Carbon abundance, $n_p$ is the proton 
number density, $g_{\rm C\,IV}$ is the fraction of Carbon that is ionized 
to \ion{C}{4}, $e$ and $m_e$ are the 
charge and mass of the electron, and $f=0.19$ is the oscillator strength.

Our ``observed optical depth'', $log_e(F/F_c)$ ($F$ is the 
observed flux and $F_c$ is the continuum level) should be an approximate 
lower bound on the Sobolev optical depth.
Except for a few complications, the Sobolev optical depth
should behave like a simple optical depth in diminishing the continuum 
flux (see
Castor 1970, Equations 7b and 20).  
First, in an idealized spherically expanding
wind there is not only ``absorption'' but also ``emission'' 
(actually forward-scattered continuum photons)--this
causes us to {\it underestimate} the true $\tau$, as desired for a lower
limit.  Second, the observed absorption arises does not arise from a 
pointlike
continuum source, and so must be averaged over impact parameter $p$. 
Equation~\ref{eq:sobolev} gives
the {\it radial} optical depth, while photons from the limb of an idealized
spherical continuum source encounter a transverse optical depth
$\tau_0=\tau_{\rm rad}(1+\sigma)$, where $\sigma=(R/v)dv/dR-1$ (McCray et 
al. 1984,
appendix), where Equation~\ref{eq:sobolev} gives $\tau_{\rm rad}$.  If the
wind velocity obeys a standard $v=v_\infty(1-R_*/R)^\beta$ law with
$\beta=0.5-1.0$, then we find that the transverse optical depth should be
within a factor of 3 of $\tau_{\rm rad}$.  With these caveats, we require 
that the Sobolev optical depth $\tau\gae1$, based on the observed profile.

We assume that $(dv/dr)^{-1} \sim
3\times10^{11}\mbox{cm}/300$~km~s$^{-1}$ and the carbon abundance
relative to hydrogen is $a_C\sim 3\times 10^{-4}$. We
also assume that the wind velocity is 300~km~s$^{-1}$ (where the
absorption is the strongest).  Although the wind is ionized by the X-ray
source it may originate on the companion star; however
we assume for simplicity that the distance from a point on the wind
to the wind's center and to the neutron star are equal, and call this $r$.
By conservation of mass we have \begin{equation}
\label{eq:conserve} n_p=\frac{\dot{M}}{\Omega r^2 v m_p}\approx
2\times10^{15} \frac{\dot{M}}{\Omega r^2} \end{equation} 
Here, $\Omega$ is the solid angle in steradians into which the wind
expands.  This should be $<4\pi$, because whether the wind
originates on the star or disk, it should not be able to expand
into the region in front of the star at $\phi=0$.
Together with
Equation~\ref{eq:sobolev} this implies $\dot{M}=2\times10^{-9}
(r^2 \Omega/4 \pi g_{\rm
C\,IV})$cm$^{-2}$~g~s$^{-1}$ and $g_{\rm C\,IV}
n_p=4\times10^{6}$~cm$^{-3}$. In order to know $g_{\rm C\,IV}$, the
fraction of C that is in the form \ion{C}{4}, we need to know the
ionization parameter $\xi\equiv L_x/r_x^2 n_p$.  We find $g_{\rm
C\,IV}=4\times10^{6} \xi r^2/L_x$. We find another relation between $\log
\xi$ and $\log g_{\rm C\,IV}$ from simulations with XSTAR, using a
power-law plus blackbody incident spectrum.  Using $r=3\times10^{11}$~cm
(a typical length scale for the wind, as this is the
separation between the stellar atmosphere and the neutron star), the two
methods give the same value of $g_{\rm C\,IV}$ at $\log \xi=2.1$, implying
$\dot{M}=2\times10^{-6} \Omega/4 \pi \Msun$~yr$^{-1}$.  This estimated
mass-loss is as
great as for O supergiants, and exceeds by three orders of magnitude the
accretion rate onto the neutron star. We then assume that the wind is not
ionized by the full X-ray lumonisty $L_x=3\times10^{37}$~erg~s$^{-1}$, but
is shadowed by, for example, the disk, so that the scattered X-rays
ionizing the wind have $L_x=3\times10^{35}$~erg~s$^{-1}$.  We then
infer
$\dot{M}=4\times10^{-8}\Omega/4\pi\Msun$~yr$^{-1}$.  While this is still
quite
substantial (an order of magnitude greater than the accretion rate), it
does not conflict with observations, in particular with the observed
orbital period change in Her~X-1 (Deeter et al. 1991).  If
$\Omega/4\pi\sim0.1$, then $\dot{M}=4\times10^{-9}\Msun$~yr$^{-1}$,
comparable to the accretion rate.

With an estimate of the wind mass-loss rate in hand, we can test the
viability of the self-excited wind model.  Although there is certainly an 
accretion disk in Hercules~X-1, accretion via a wind may still play a role.  
For example, Friend \&\ Castor (1982) and Blondin, Stevens, \&\ Kallman 
(1991) have shown that as a mass-donor star approaches its Roche lobe, there 
is a continuous transition between the classical Bondi-Hoyle wind accretion 
and Roche lobe gas stream mass transfer.

Assuming that the wind detected in the UV lines is
driven from the X-ray heated atmosphere of HZ~Her, what fraction of the
X-ray luminosity of Her~X-1 could be powered by Bondi-Hoyle capture of the
gas?  The wind-accretion luminosity is given by
\begin{equation}
L_{\rm X}=\pi \zeta r_{\rm acc}^2 v_{\rm rel} \rho_{\rm ns} \frac{G M_{\rm
ns}}{R_{\rm ns}}
\end{equation}
Here, $\zeta\sim0.1$ is the energy efficiency for accretion onto a neutron
star, $M_{\rm ns}$ and $R_{\rm ns}$ are the mass and radius of the neutron
star, $\rho_{\rm ns}$ is the
density of the stellar wind near the neutron star, and $v_{\rm
rel}=(v_{\rm
wind}^2+v_{\rm ns})^{1/2}$ is the velocity of the wind (with
radial velocity $v_{\rm wind}$) relative to the
neutron star (with orbital velocity $v_{\rm ns}$).  The accretion
capture radius $r_{\rm acc}$ is determined simply from the escape velocity
of the neutron star by
\begin{equation}
r_{\rm acc}=\frac{2 G M_{\rm ns}}{v_{\rm rel}^2}
\end{equation}
We use standard values of $M_{\rm ns}=1.4\Msun$, $R_{\rm ns}=10$~km,
and $v_{\rm ns}=169$~km~s$^{-1}$.  We find $\rho_{\rm ns}$ from $\rho_{\rm
ns}=m_p n_p$ at the neutron star, assumed to be a distance
$r=3\times10^{11}$~cm from the source of the wind in this case. We take
$\dot{M}=4\times10^{-8}
\Omega/4 \pi \Msun$~yr$^{-1}$, as found from the UV lines.  

The velocity of the wind at the neutron star's orbit is not necessarily
the wind's terminal velocity.  There is thus a large uncertainty in the
value of $v_{\rm rel}$, and this has a large effect on our answer for
$L_{\rm X}$.  For $v_{\rm rel}=300$~km~s$^{-1}$ we find $L_{\rm
X}=2\times10^{37}$~erg~s$^{-1}$, comparable to the observed X-ray
luminosity, while for $v_{\rm rel}=400$~km~s$^{-1}$, $L_{\rm
X}=7\times10^{36}$~erg~s$^{-1}$.  Thus we conclude that {\it if} this wind
originates in the atmosphere of HZ~Her and not on the accretion disk, that
it is possible for accretion of this wind to power a substantial portion
or even all of the X-ray output of the neutron star.

We investigate the hypothesis that the emission at $\phi=0$, the
absorption at $\phi=0.1-0.3$, and the P~Cygni absorption seen at
$\phi=0.10$, $\phi=0.3-0.4$, $\phi=0.25-0.75$ are due to the same gas. The
main argument
for this hypothesis is that the narrow absorption lines (Boroson et al.
2000a and Figures~1,2) are stationary over a large fraction of the orbit.  
In both the 1998 and 1999 data sets, we find no detectable motion in the
narrow line from $\phi=0.1$ to $\phi=0.3$.  During this interval the
velocity of the neutron star has shifted by 60~km~s$^{-1}$.  Material on
HZ~Her would not seem a likely candidate for the absorption line gas as at
these phases, as it would not cover the broad line and continuum emission,
which we attribute to the disk.  Thus we expect that the absorbing gas, in
common with the gas that provides the emission lines at $\phi=0$, fills a
region larger than the binary orbit.


Much of the flux in the resonance lines seen in mid-eclipse may be due
to photons scattered in a wind instead of true emission.
If all photons that are
``absorbed'' by the narrow lines are merely scattered out of our line
of sight, then we expect to
see a flux of scattered emission equal to the orbit-averaged absorbed
flux. This would be simply the equivalent width of the line (say 0.2\AA\
for \ion{N}{5}$\lambda1238.8$) times the average flux at 1237\AA\
(blue-shifted from the rest wavelength by $\approx400$~km~s$^{-1}$).  The
flux available for scattering should include the continuum flux as well as
the broad line emission from the accretion disk, which should extend to
blue-shifts of $\approx300-400$~km~s$^{-1}$.  A reasonable estimate for
this flux, from Figure~1 and Boroson et al. (2000a), is
$5\times10^{-14}$~erg~s$^{-1}$~cm$^{-2}$~\AA$^{-1}$.  (This is valid for a
first approximation only, as the equivalent widths of
the absorption lines are variable, we did not detect them from
$\phi=0.7-0.8$.)  The total absorbed flux is then
$10^{-14}$~erg~s$^{-1}$~cm$^{-2}$.  For the 1242.8\AA\ line, which has a
lower oscillator strength and absorption lines with lower equivalent
width, we would expect a flux of
$\approx7\times10^{-15}$~erg~s$^{-1}$~cm$^{-2}$.  This estimate is within
50\%\ of the observed values of the emission line flux seen in
mid-eclipse (Anderson et al. 1994): $1.5,
1.2\times10^{-14}$~erg~s$^{-1}$~cm$^{-2}$.

The problem with a scattering origin for the emission lines seen in
mid-eclipse is that the \ion{N}{4}$\lambda1486$ and
\ion{He}{2}$\lambda1640$ lines do not show either narrow or P~Cygni
absorption and from reasons of atomic physics are not
expected to scatter photons.  For example, the \ion{N}{4}$\lambda1486$ is
semi-forbidden and has an oscillator strength $f=5.74\times10^{-6}$
(compare with \ion{C}{4}$\lambda1548.195$ which has $f=0.19$, Verner et
al. 1994).  

If a substantial portion of the flux in the resonance lines is the
result of scattering, then the {\it true} emission by the wind must have a
higher fraction of \ion{N}{4}$\lambda1486$ and \ion{He}{2}$\lambda1640$
relative to the resonance lines than the broad and narrow line regions.  
When we simulated the line emission with XSTAR using values for density
and
ionization stage of the emitting gas similar to those of Anderson et al.,
we could not obtain a UV emission spectrum enhanced in both
\ion{N}{4}$\lambda1486$ and \ion{He}{2}$\lambda1640$ emission.  For
$\log\xi\lae 1.5$, \ion{N}{4} becomes stronger (and \ion{He}{2} becomes
weaker), while for $\log\xi\gae 1.5$, \ion{He}{2} becomes stronger (and
\ion{N}{4} becomes weaker).

There are other scenarios that can explain the \ion{He}{2} emission seen
in mid-eclipse.  For example, a strong continuum flux at
\ion{He}{2}Ly$\beta$ (at 256\AA) could raise He to the n$=3$ level and
result in $\lambda1640$ emission.  Using the model given by Dal Fiume
et al. (1998) for the soft X-ray
blackbody component to the Her~X-1 spectrum, we estimate flux at 256\AA\
of $2.5\times10^{-14}$~erg~s$^{-1}$~cm$^{-2}$~\aa$^{-1}$.  Given the
far greater cosmic abundance of He relative to N, and that the oscillator
strengths of \ion{He}{2}Ly$\beta$ and \ion{N}{5}$\lambda1238.8,1242.8$ are
of the same order of magnitude, it does not seem unreasonable for
\ion{He}{2}Ly$\beta$ to have an equivalent width of $\sim1$\AA, which
could produce the observed \ion{He}{2}$\lambda1640$ emission flux of
$1.2,1.1\times10^{-14}$~erg~s$^{-1}$~cm$^{-2}$
in mid-eclipse.

Furthermore, optical \ion{He}{2}$\lambda4686$ emission is seen in
mid-eclipse
(Anderson et al. 1994; Still et al. 1997).  A similar emission mechanism
may be responsible for this line, and in fact, each \ion{He}{2} electron
that goes from
n=4 to n=3 (emitting a 4686\AA\ photon) may then go down to the n=2 level
and emit a 1640 \AA\ photon.  From the observed flux of
\ion{He}{2}$\lambda4686$, $5\times10^{-15}$~erg~s$^{-1}$~cm$^{-2}$, we
would then infer a \ion{He}{2}$\lambda1640$ flux of
$1.4\times10^{-14}$~erg~s$^{-1}$~cm$^{-2}$.  

Thus it is possible that the permitted resonance lines seen in mid-eclipse
are actually scattered, in which case we would expect them to be
red-shifted.
Scattering can only occur from
surrounding regions that have a direct view of parts
of the system that are bright in the UV (the disk and heated face of the
star).  If the gas is blue-shifted when it is in front of the emission,
causing the narrow absorption lines, it should be red-shifted at $\phi=0$.
If the \ion{He}{2} line is formed by radiative excitation to the n=3 or
higher levels, then it should be emitted only by portions of the wind in
view of the soft X-ray source, and thus should also be red-shifted in
mid-eclipse.

Although Anderson et al. suggested tentatively that the lines seen in
mid-eclipse
were redshifted, the redshift was comparable to the accuracy of the FOS
wavelengths.  The line profiles at $\phi=0.057, 0.065$ observed with the
STIS (O4V403010 and O4V406010) show both a
blue-shifted component and a weaker red-shifted component (extending to
$\approx400$~km~s$^{-1}$ in \ion{N}{5} and \ion{C}{4}. The blue-shifted
component may arise from the receding part of the disk (the first to
emerge from mid-eclipse), and the red-shifted component may be the same as
that seen in mid-eclipse, as the line fluxes are similar.  
We have fit the lines seen in mid-eclipse with the STIS in low resolution
with Gaussians.  The lines are near rest velocity or slightly blueshifted
($\sim100$~km~s$^{-1}$), but there are large systematic errors associated
with fitting doublets with absorption lines that may overlap the emission.
Further observations of the lines seen in mid-eclipse should clarify the
contribution of scattering to the emission.

The P~Cygni lines are variable, appearing only
at $\phi\approx0.25-0.75$, and
only in
absorption at $\phi=0.1$.  Furthermore, these features do not always
appear at these phases, and may be absent outside of the anomalous low
state of 1999. Are they variable because the outflow starts and
stops, or because the illumination of the wind by X-rays is variable?
The previous calculation shows that a more reasonable value of $\dot{M}$
follows if we assume that the P~Cygni profile can only form where the wind
is shadowed from direct X-ray illumination.  Furthermore, the absorption
at $\phi=0.10$ disappears by $\phi=0.14$.  With an observed maximum 
line-of-sight wind
velocity of 250~km~s$^{-1}$, the leading edge of the wind would travel
by just $2\times10^{11}$~cm, the radius of the accretion disk, between
the end of one observation and the start of the next.  Thus it seems more
likely that the absorbing material has been ionized than that the flow
has become unsteady.

The similarity in velocities between the narrow absorption lines and the
broader P~Cygni absorption features suggests a physical link.  Perhaps the
narrow absorption lines arise in gas that is coasting after having
already been accelerated in the region that causes the P~Cygni lines.

\section{Future Work}

We are experimenting with hydrodynamic models of a wind in Hercules~X-1.
From the output of these models, we will predict the UV emission and
absorption line profiles as a function of orbital phase.  We hope that
this will help us distinguish the observational signatures of winds from
the accretion disk and winds from HZ~Her.

The narrow resonance emission lines may be emitted or scattered in a wind.
Examination of the velocities of gaussian fits to the narrow lines
(Boroson et al. 2000a) shows that they do not match the
expected velocity of the X-ray heated, visible face of HZ~Her
(Boroson et al. 1996).  Doppler tomograms of the narrow lines (Vrtilek et
al. 2001) place them on one side of the face of HZ~Her; however with
this placement the lines would be brighter at $\phi=0.25$ than at
$\phi=0.75$.  The observed lines behave in exactly
the opposite manner,
and are $\sim4$ times brighter at $\phi=0.75$.  Non-resonance lines,
less likely to be formed by scattering, show tomograms more evenly
spread throughout the Roche lobe of HZ~Her.  One possible explanation
for the difficulty in matching the narrow line velocity with a model is 
that the
velocity of the narrow lines is not determined simply from the system's
1.7~day rotation.  We will investigate the possibility that the narrow
line profile is affected by the motion of a wind in a future paper.

\section{Acknowledgements}

Based on observations with the NASA/ESA {\it Hubble Space Telescope},
obtained at the Space Telescope Science Institute, which is operated by
the Association of Universities for Research in Astronomy, Inc., under
NASA contract GO-05874.01-94A.  SDV supported in part by NASA
(NAG5-2532, NAGW-2685), and NSF (DGE-9350074).  BB acknowledges an NRC
postdoctoral associateship.

\clearpage

\figcaption{Top panel: FOS spectrum of Her~X-1 from a 1600~s exposure
spanning $\phi=0.995-0.006$.  The strong emission line at 1216\AA\
is geocoronal, and not intrinsic to Her~X-1.  Bottom panel: a 444~s STIS
exposure spanning $\phi=0.984-0.990$.}

\figcaption{Narrow absorption lines in
\ion{N}{5}$\lambda\lambda1238.8,1242.8$.  Vertical dotted lines mark
a heliocentric velocity of $-400$~km~s$^{-1}$.  (a) Observations
with abbreviated root name 80x0 (solid line, see Table~2 for 
observation root 
names), with average phase $\phi=0.27$, and
with root name 60x0 (dot-dashed line), with
$\phi=0.14$.  (b) Observations with root name 70x0 (solid line),
with $\phi=0.70$, and with root name 50x0 (dot-dashed
line), with $\phi=0.59$.}

\figcaption{Narrow absorption lines in \ion{C}{4}$\lambda\lambda
1548.195,1550.77$.  There are strong interstellar absorption lines near
the rest wavelengths of the \ion{C}{4} lines.  Vertical lines mark a
heliocentric velocity of $-400$~km~s$^{-1}$.  (a) Observations with root
name 80x0 (solid line), with average phase $\phi=0.27$, and with root name
60x0 (dot-dashed line), with $\phi=0.14$.  (b) Observations with root name
70x0 (solid line), with $\phi=0.70$, and with root name 50x0 (dot-dashed
line), with $\phi=0.59$.}

\figcaption{P~Cygni line profiles in the UV spectrum of Hercules~X-1.
We have averaged the orbits with root name O4V50x0 and 7010, 7020, 7030.
(a) \ion{C}{4}$\lambda\lambda 1548,1550$ (strong ISM absorption lines
appear at the rest wavelengths of the lines), (b)
\ion{Si}{4}$\lambda\lambda
1393,1403$ (and a blend of \ion{O}{4} and \ion{Si}{4} lines), and
(c) \ion{N}{5}$\lambda\lambda 1238.8,1242.8$.}

\figcaption{A comparison of the line profiles at $\phi=0.10$ (solid line)
and $\phi=0.14$ (dot-dashed line), showing absorption at velocities from
$v=-250$~km~s$^{-1}$ to $v=0$.  (a)
N\,V$\lambda\lambda1239,1243$, (b) O\,V$\lambda1371$, (c)
Si\,IV$\lambda\lambda1393,1403$, (d)
C\,IV$\lambda\lambda1548,1551$, (e) He\,II$\lambda1640$.
The top x-axis marks the heliocentric velocity; for the doublets we
use the blue component.}

\figcaption{We sketch the components of the Her~X-1 system, including
HZ~Her, the neutron star (marked with X), the accretion disk, and the gas
stream.  The UV continuum is emitted mostly from the heated face of
HZ~Her, as shown.  At $\phi=0.5$, the UV emission is absorbed by
approaching gas, and is blue-shifted.  P~Cygni absorption is caused by gas
close to the system that is still accelerating, while narrow absorption is
caused by gas further out that coasts at the wind's terminal velocity.}

\clearpage

\begin{deluxetable}{ccccc}
\tablecaption{HST Far UV Observations of Hercules~X-1}
\tablehead{
\colhead{Root name}  & \colhead{Instrument\tablenotemark{a}} & 
\colhead{Start (MJD)\tablenotemark{b}} &
\colhead{Exposure (s)} & 
\colhead{Orbital Phase\tablenotemark{c}}}
\startdata
Y0XI0201 & FOS H & 48742.573 & 1600 &  0.458\\
Y0XI0204 & FOS H & 48748.595 & 1600 &  0.961\\
\tableline
Z2HP0202  & GHRS L & 49591.134  & 1266 &  0.559\\
Z2HP0204  & GHRS L & 49591.199  & 729 &  0.596\\
Z2HP0206  & GHRS M & 49591.267  & 2311  & 0.641\\
Z2HP0208  & GHRS M & 49591.333  & 2474  & 0.680\\
Z2HP020A  & GHRS L & 49591.400  & 911  & 0.715\\
Z2HP020C  & GHRS L & 49591.413  & 963  & 0.723\\
Z2HP020E  & GHRS M & 49591.467  & 2431  & 0.760\\
Z2HP020G  & GHRS M & 49591.534  & 2475  & 0.799\\
Z2HP020I  & GHRS L & 49591.601  & 911  & 0.833\\
Z2HP020K  & GHRS L & 49591.614  & 969  & 0.841\\
Z2HP020M  & GHRS M & 49591.669  & 2413  & 0.878\\
Z2HP020O  & GHRS M & 49591.735  & 1585  & 0.914\\
\tableline
O4V401010 & STIS L & 51004.563 & 565  & 0.904\\  
O4V401020 & STIS L & 51004.572 & 1517  & 0.913\\
O4V401030 & STIS L & 51004.623 & 1495  & 0.944\\
O4V401040 & STIS L & 51004.646 & 1702  & 0.967\\ 
O4V401050 & STIS L & 51004.702 & 444 &  0.985\\
O4V403010 & STIS E & 51006.514 & 2227  & 0.057\\
O4V403020 & STIS E & 51006.572 & 2636 & 0.092\\
O4V403030 & STIS E & 51006.639 & 2636  & 0.132\\
O4V403040 & STIS E & 51006.706 & 2636   & 0.171\\
O4V403050 & STIS E & 51006.773 & 2620  & 0.211\\
O4V404010 & STIS E & 51007.522 & 2227  & 0.650\\
O4V404020 & STIS E & 51007.580 & 2636 & 0.685\\
O4V404030 & STIS E & 51007.647 & 2636 & 0.725\\
O4V404040 & STIS E & 51007.714 & 2636 & 0.764\\
O4V404050 & STIS E & 51007.781 & 2620 & 0.804\\
\tableline
\tablebreak
O4V405010 & STIS E & 51371.124 & 2227 & 0.512\\ 
O4V405020 & STIS E & 51371.185 & 2636 & 0.549\\
O4V405030 & STIS E & 51371.253 & 2636 & 0.589\\
O4V405040 & STIS E & 51371.321 & 2636 & 0.629\\
O4V405050 & STIS E & 51371.388 & 2620 & 0.669 \\
O4V406010 & STIS E & 51372.064 & 2227 & 0.065 \\
O4V406020 & STIS E & 51372.124 & 2636 & 0.102 \\
O4V406030 & STIS E & 51372.193 & 2636 & 0.142 \\
O4V406040 & STIS E & 51372.260 & 2636 & 0.182 \\
O4V406050 & STIS E & 51372.328 & 2620 & 0.221 \\
O4V407010 & STIS E & 51373.004 & 2227 & 0.618 \\
O4V407020 & STIS E & 51373.065 & 2636 & 0.655 \\
O4V407030 & STIS E & 51373.133 & 2636 & 0.695 \\
O4V407040 & STIS E & 51373.200 & 2636 & 0.735 \\
O4V407050 & STIS E & 51373.268 & 2620 & 0.774 \\
O4V408010 & STIS E & 51374.011 & 2227 &  0.210 \\
O4V408020 & STIS E & 51374.072 & 2636 & 0.247 \\
O4V408030 & STIS E & 51384.140 & 2636 & 0.288 \\
O4V408040 & STIS E & 51374.207 & 2620 & 0.327 \\
\enddata
\tablenotetext{a}{L means low resolution, M means medium
resolution, H means high resolution, and E means echelle mode}
\tablenotetext{b}{In terms of calendar day, the FOS observations took 
place April 30 and May 6, 1992, the GHRS observations took place August 
27, 1994, 
the STIS observations
O4V401010 through 4050 took place from July 10-15, 1998, and STIS 
observations O4V405010 through O4V408030 took place from July 12-15, 
1999.}

\tablenotetext{c}{The orbital phase of the mid-exposure time, using the
ephemeris of Deeter et al. (1991)}
\end{deluxetable}

\begin{deluxetable}{lccccc}
\tablecaption{Hercules X-1 Absorption Lines}
\tablehead{
\colhead{Root name}  & \colhead{Phase\tablenotemark{a}} & 
\colhead{Line\tablenotemark{b}} &
\colhead{$W_\lambda$} & \colhead{Type\tablenotemark{c}} & \colhead{V
range\tablenotemark{d}}} 
\startdata
O4V403020 & 0.092 & \ion{N}{5}B & 0.15$\pm0.01$ & N & 
-513$\pm1$, (-572,-470)$\pm(1,10)$\\
O4V403020 & 0.092 & \ion{N}{5}R & $0.094\pm0.006$ & N &
-503$\pm2$, (-550,-460)$\pm(20,10)$\\
O4V403030 & 0.132 & \ion{N}{5}B & 0.085$\pm0.006$ & N &
-510$\pm2$, (-550,-460)$\pm(20,10)$\\
O4V403030 & 0.132 & \ion{N}{5}R & 0.060$\pm0.009$ & N &
-500$\pm1$, (-540,-450)$\pm(10,10)$\\
O4V403040 & 0.171 & \ion{N}{5}B & 0.091$\pm0.03$ & N &
-494$\pm5$, (-560,-430)$\pm(10,10)$\\ 
O4V403040 & 0.171 & \ion{N}{5}R & 0.064$\pm0.005$ & N &
-515$\pm2$, (-560,-490)$\pm(20,20)$\\
\tableline
O4V405010 & 0.512 & \ion{N}{5}B & $0.60\pm0.04$ & P &
(-520,-200)$\pm(5,20)$\\
O4V405010 & 0.512 & \ion{Si}{4}B & $0.53\pm0.01$ & P & 
(-520,-100)$\pm(10,10)$\\
O4V405010 & 0.512 & \ion{C}{4}B & $0.79\pm0.06$ & P & 
(-600,-160)$\pm(30,10)$\\
O4V405020 & 0.549 & \ion{N}{5}B & $0.52\pm0.01$ & P &
(-550,-260)$\pm(10,10)$\\
O4V405020 & 0.549 & \ion{Si}{4}B & $0.25\pm0.02$ & P &
(-520,-160)$\pm(10,10)$\\
O4V405020 & 0.549 & \ion{C}{4}B & $0.58\pm0.07$ & P & 
(-530,-210)$\pm(30,10)$\\
O4V405030 & 0.589 & \ion{N}{5}B & $0.34\pm0.03$ & P &
(-520,-270)$\pm(40,10)$\\
O4V405030 & 0.589 & \ion{Si}{4}B & $0.17\pm0.02$ & P &
(-560,-200)$\pm(10,40)$\\
O4V405030 & 0.589 & \ion{C}{4}B & $0.57\pm0.05$ & P &
(-500,-220)$\pm(10,10)$\\
O4V405040 & 0.629 & \ion{N}{5}B & $0.30\pm0.04$ & P &
(-530,-280)$\pm(10,10)$\\
O4V405040 & 0.629 & \ion{Si}{4}B & 0.09$\pm0.05$ & P &
(-590,-190)$\pm(60,10)$\\ 
O4V405040 & 0.629 & \ion{C}{4}B & $0.37\pm0.03$ & P & 
(-480,-260)$\pm(10,10)$\\
O4V405050 & 0.669 & \ion{N}{5}B & $0.28\pm0.02$ & P &
(-530,-290)$\pm(100,10)$\\
O4V405050 & 0.669 & \ion{Si}{4}B & $0.08\pm0.01$ & P &
(-510,-340)$\pm(30,10)$\\
O4V405050 & 0.669 & \ion{C}{4}B & $0.31\pm0.05$ & P &
(-450,-280)$\pm(10,10)$\\
O4V406020 & 0.102 & \ion{N}{5}B & $0.71\pm0.01$ & P? &
(-280,30)$\pm(10,10)$\\
O4V406020 & 0.102 & \ion{N}{5}R & $0.56\pm0.04$ & P? &
(-260,0)$\pm(10,10)$\\
O4V406020 & 0.102 & \ion{N}{5}B & $0.21\pm0.02$ & N & 
-394$\pm1$, (-440,-350)$\pm(10,10)$\\
O4V406020 & 0.102 & \ion{C}{4}B & $0.20\pm0.01$ & N & 
-380$\pm1$, (-430,-340)$\pm(10,10)$\\
O4V406020 & 0.102 & \ion{C}{4}R & $0.17\pm0.02$ & N &
-394$\pm1$, (-430,-360)$\pm(10,10)$\\
O4V406030 & 0.142 & \ion{N}{5}B & $0.22\pm0.02$ & N &
-393$\pm3$, (-440,-340)$\pm(10,10)$\\
O4V406030 & 0.142 & \ion{C}{4}B & $0.23\pm0.01$ & N &
-394$\pm1$, (-440,-350)$\pm(10,10)$\\
O4V406030 & 0.142 & \ion{C}{4}R & $0.17\pm0.01$ & N &
-400$\pm1$, (-440,-350)$\pm(10,10)$\\
O4V406040 & 0.182 & \ion{N}{5}B & $0.29\pm0.03$ & N &
-409$\pm4$, (-480,-340)$\pm(10,10)$\\
O4V406040 & 0.182 & \ion{C}{4}B & $0.150\pm0.004$ & N &
-410$\pm1$, (-440,-380)$\pm(10,10)$\\
O4V406040 & 0.182 & \ion{C}{4}R & $0.075\pm0.01$ & N & 
-409$\pm2$, (-440,-380)$\pm(10,10)$\\
O4V406050 & 0.221 & \ion{N}{5}B & $0.42\pm0.02$ & N,P? &
-447$\pm2$, (-550,-350)$\pm(10,10)$\\ 
O4V406050 & 0.221 & \ion{C}{4}B & $0.22\pm0.01$ & N & 
-425$\pm3$, (-480,-380)$\pm(10,10)$\\
O4V406050 & 0.221 & \ion{C}{4}R & $0.08\pm0.02$ & N &
-426$\pm3$, (-450,-390)$\pm(10,10)$\\
O4V407010 & 0.618 & \ion{N}{5}B & $0.22\pm0.03$ & P & 
(-590,-290)$\pm(20,10)$\\
O4V407010 & 0.618 & \ion{Si}{4}B & $0.17\pm0.03$ & P & 
(-560,-170)$\pm(30,10)$\\
O4V407010 & 0.618 & \ion{C}{4}B & $0.38\pm0.02$ & P & 
(-480,-260)$\pm(10,10)$\\
O4V407020 & 0.655 & \ion{N}{5}B & $0.31\pm0.02$ & P &
(-510,-290)$\pm(20,10)$\\
O4V407020 & 0.655 & \ion{Si}{4}B & $0.20\pm0.04$ & P & 
(-500,-170)$\pm(20,10)$\\
O4V407020 & 0.655 & \ion{C}{4}B & $0.45\pm0.05$ & P &
(-480,-260)$\pm(20,10)$\\
O4V407030 & 0.695 & \ion{N}{5}B & $0.35\pm0.06$ & P &
(-470,-290)$\pm(40,10)$\\
O4V407030 & 0.695 & \ion{C}{4}B & $0.41\pm0.02$ & P &
(-460,-270)$\pm(30,10)$\\
O4V407040 & 0.735 & \ion{N}{5}B & $0.36\pm0.04$ & P &
(-530,-300)$\pm(20,10)$\\
O4V407040 & 0.735 & \ion{C}{4}B & $0.37\pm0.02$ & P &
(-440,-290)$\pm(10,10)$\\
O4V407050 & 0.774 & \ion{C}{4}B & $0.37\pm0.02$ & P &
(-400,-290)$\pm(10,10)$\\
\tableline
\tablebreak
O4V408010 & 0.210 & \ion{N}{5}B & $0.33\pm0.01$ & N &
$-399\pm1$(-460,-340)$\pm(10,10)$\\
O4V408020 & 0.247 & \ion{N}{5}B & $0.33\pm0.21$ & P &
(-490,-220)$\pm(170,170)$\\
O4V408020 & 0.247 & \ion{N}{5}R & $0.09\pm0.01$ & N &
$-397\pm2$(-420,-350)$\pm(10,10)$\\
O4V408020 & 0.247 & \ion{Si}{4}B & $0.42\pm0.09$ & P &
(-390,-150)$\pm(70,10)$\\
O4V408020 & 0.247 & \ion{C}{4}B & $0.24\pm0.01$ & P &
(-280,-170)$\pm(10,10)$\\
O4V408030 & 0.288 & \ion{N}{5}B & $0.48\pm0.06$ & P &
(-550,-150)$\pm(10,10)$\\
O4V408030 & 0.288 & \ion{N}{5}R & $0.20\pm0.03$ & N &
$392\pm1$,(-450,-330)$\pm(10,10)$\\
O4V408030 & 0.288 & \ion{Si}{4}B & $0.62\pm0.02$ & P &
(-470,-120)$\pm(10,10)$\\
O4V408030 & 0.288 & \ion{C}{4}B & $0.61\pm0.03$ & P &
(-490,-150)$\pm(10,10)$\\
O4V408040 & 0.327 & \ion{N}{5}B & $0.53\pm0.02$ & P &
(-500,-160)$\pm(20,10)$\\
O4V408040 & 0.327 & \ion{N}{5}B & $0.20\pm0.03$ & N & 
$-402\pm4$,(-460,-360)$\pm$(20,10)\\
O4V408040 & 0.327 & \ion{N}{5}R & $0.09\pm0.03$ & N &
$-404\pm4$,(-440,-360)$\pm(20,20)$\\
O4V408040 & 0.327 & \ion{Si}{4}B & $0.33\pm0.02$ & P &
(-450,-130)$\pm(10,10)$\\
O4V408040 & 0.327 & \ion{C}{4}B & $0.58\pm0.03$ & P &
(-460,-140)$\pm(20,10)$\\
\enddata
\tablenotetext{a}{The orbital phase of the mid-exposure time, using the
ephemeris of Deeter et al. (1991)}
\tablenotetext{b}{B indicates blue doublet component and R indicates red
doublet component}
\tablenotetext{c}{P=P Cygni line absorption; P?=possible P Cygni
absorption, N=narrow absorption line}
\tablenotetext{d}{For P Cygni absorption, we give the the range of
heliocentric Doppler velocities (in km~s$^{-1}$), while for the narrow
absorption lines, we also give the centroid velocity.}
\end{deluxetable}

\clearpage

\setcounter{figure}{0}

\begin{figure}
\caption{}
\plotone{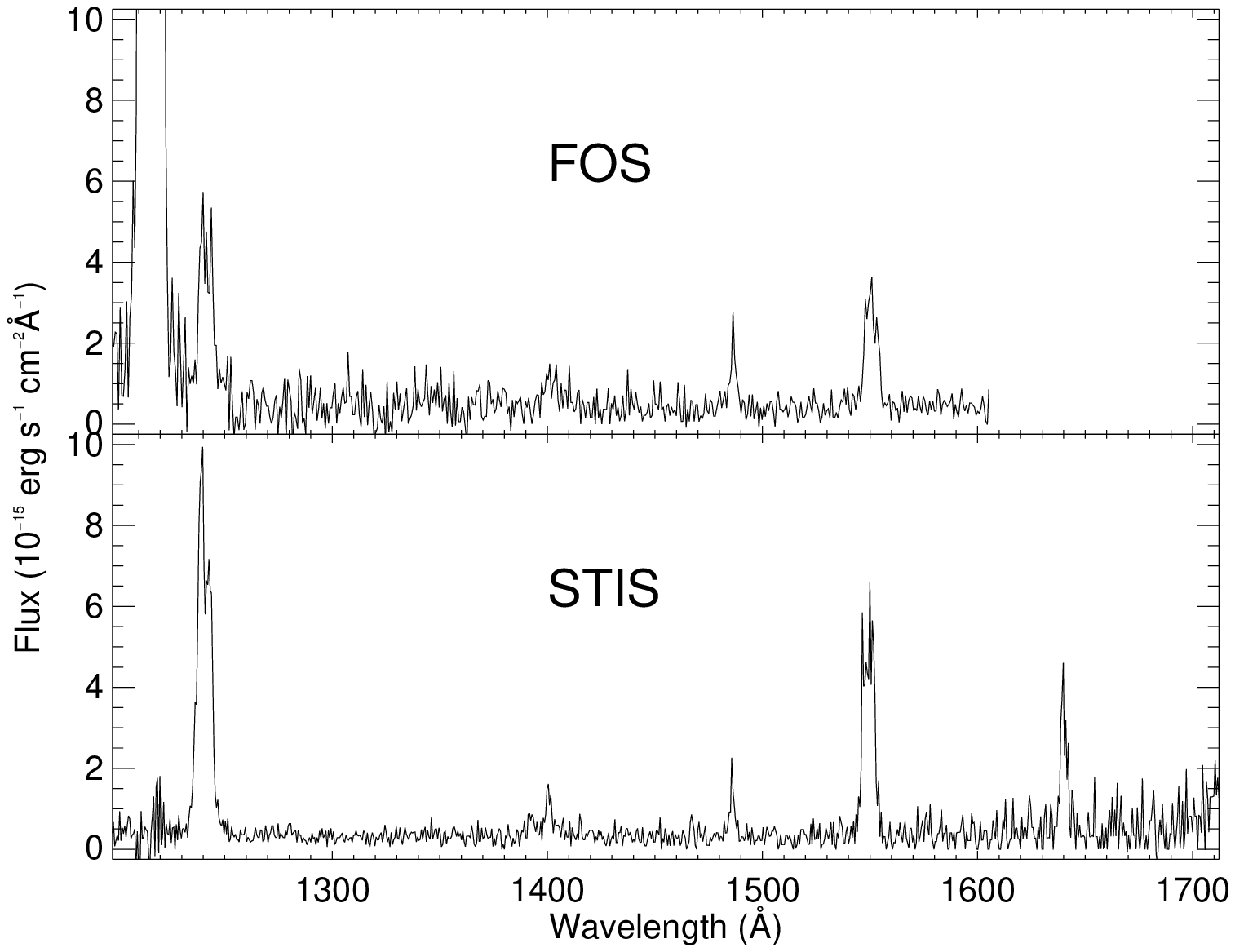}
\end{figure}

\begin{figure}
\caption{}
\plotone{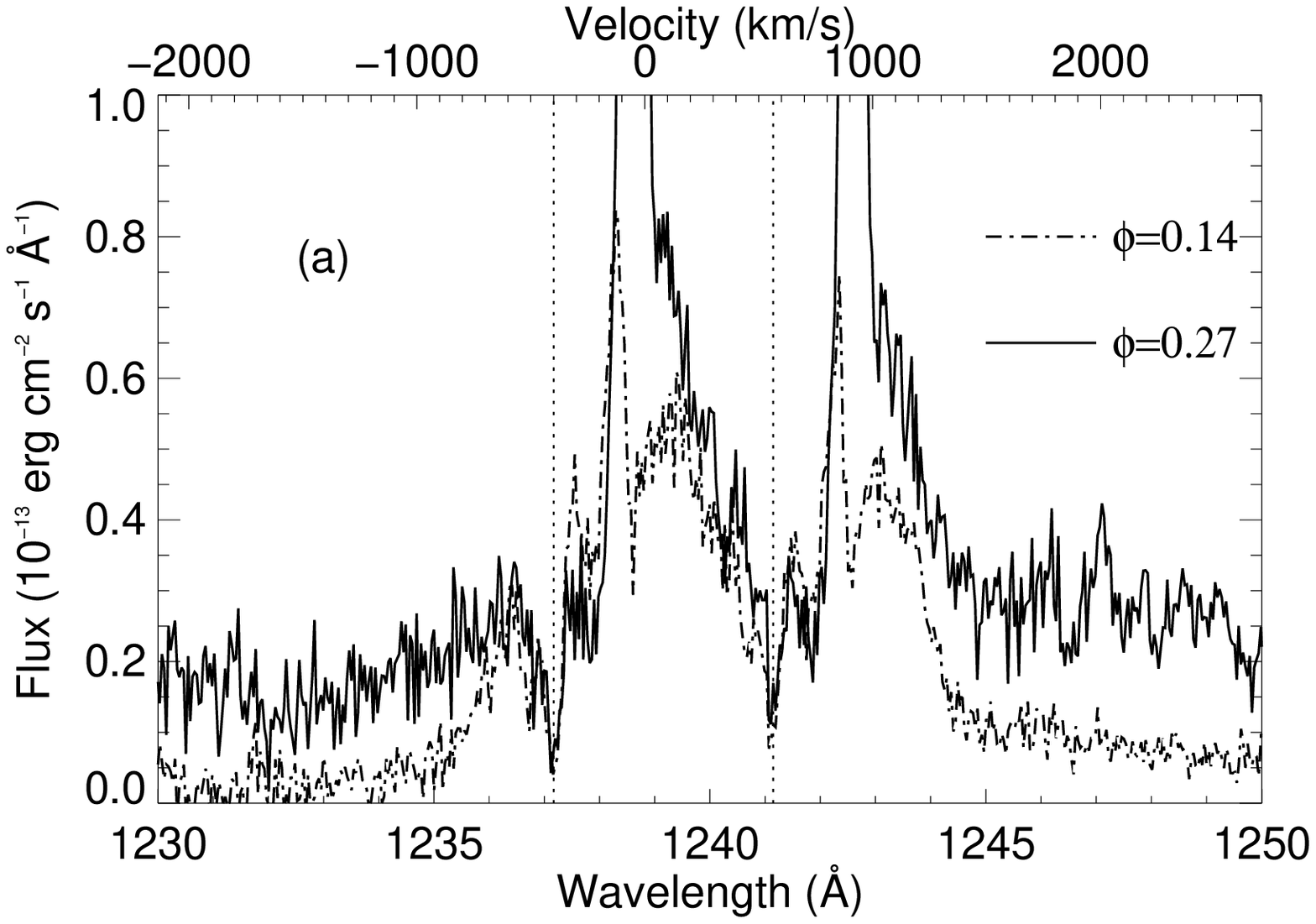}\\
\plotone{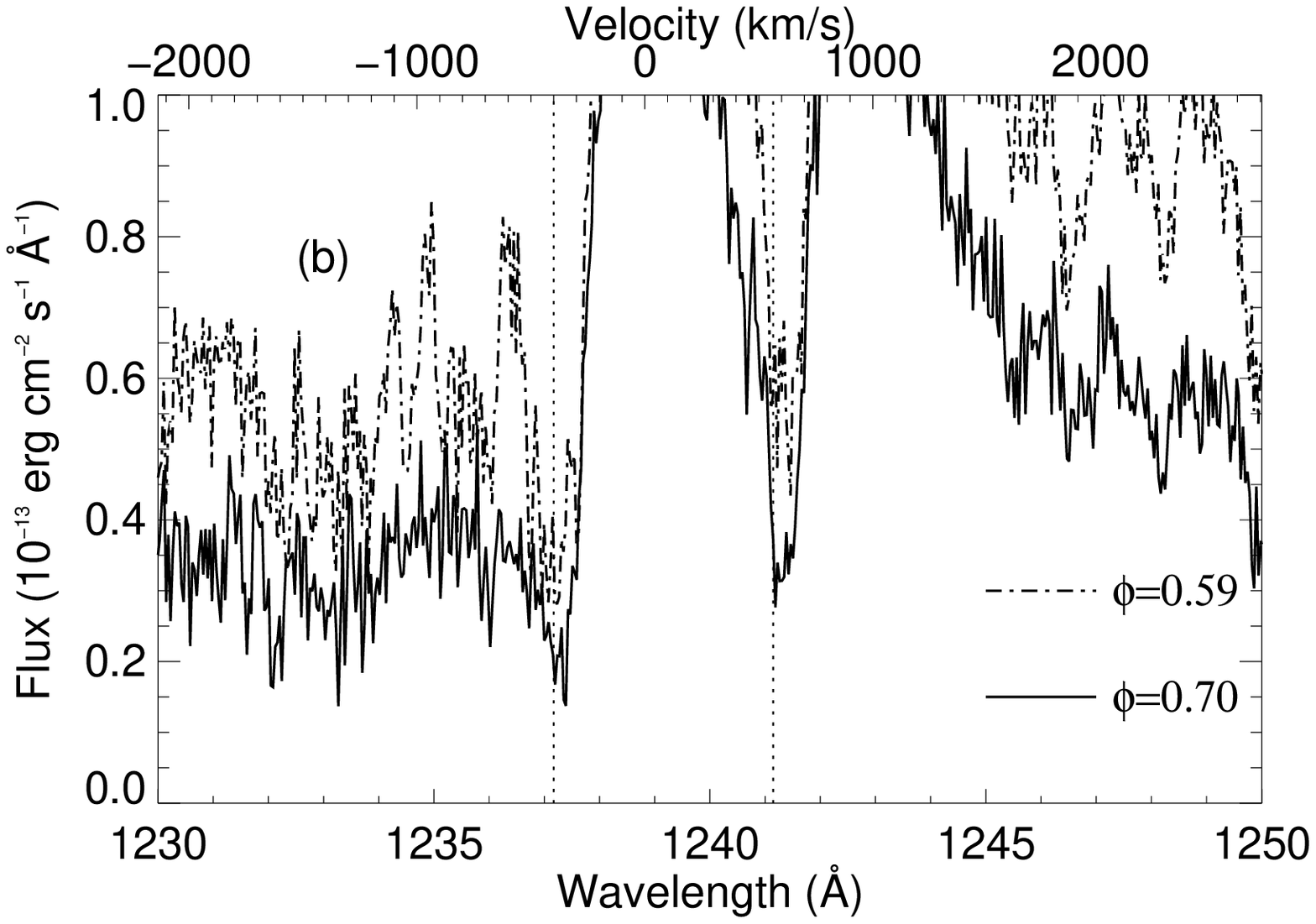}
\end{figure}

\setcounter{figure}{2}

\begin{figure}
\caption{}
\plotone{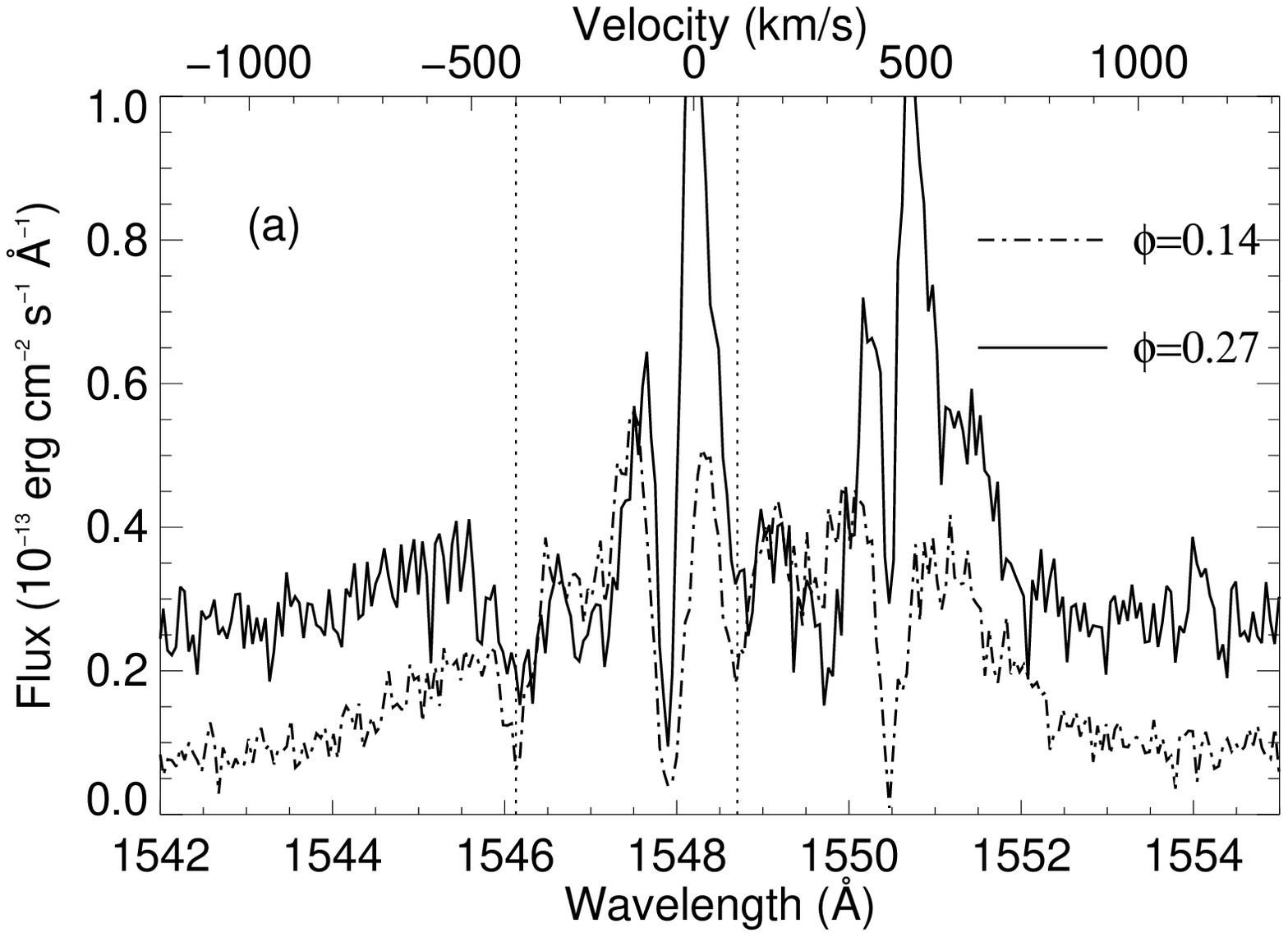}\\
\plotone{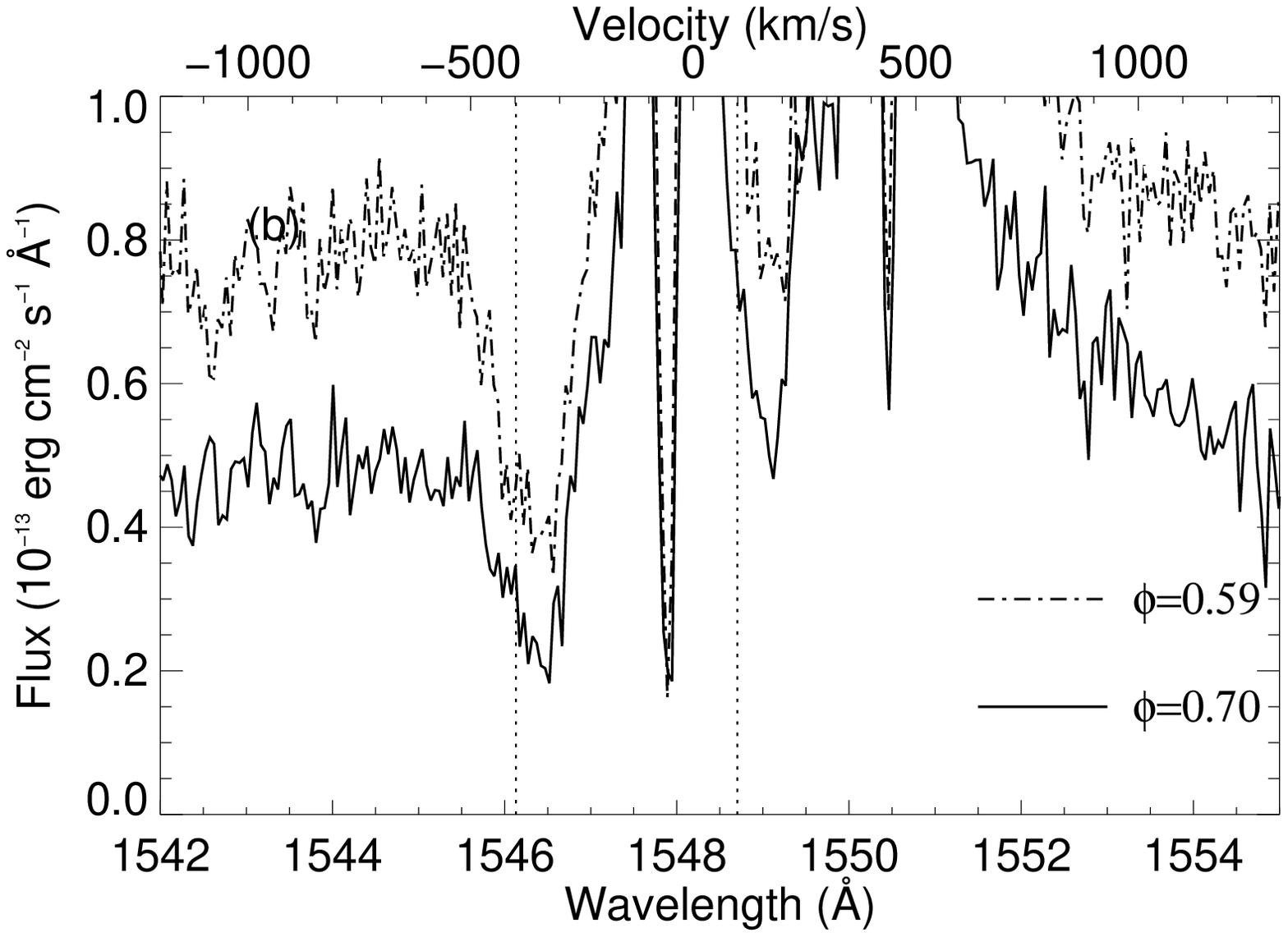}
\end{figure}

\begin{figure}
\caption{}
\plotone{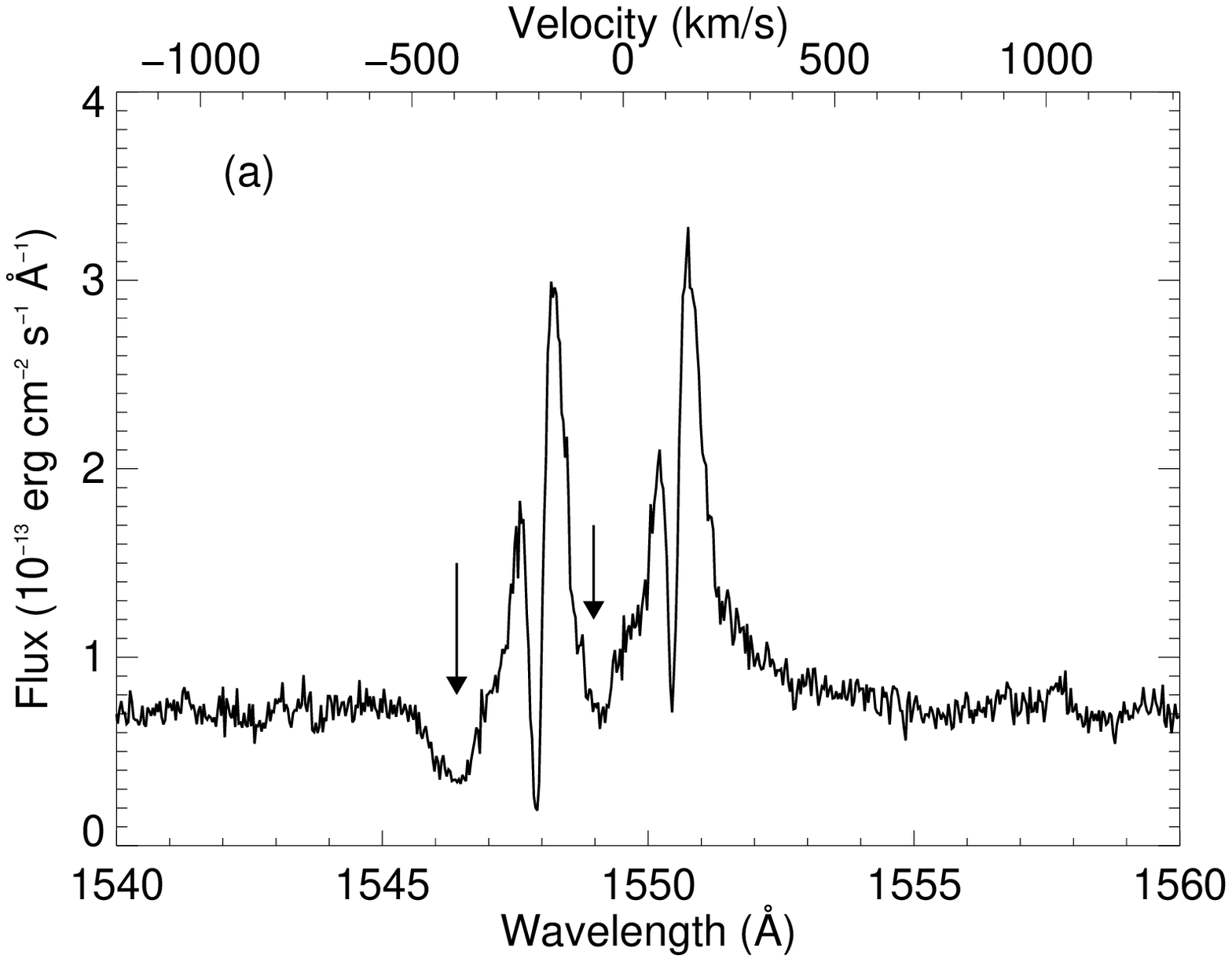}\\
\plotone{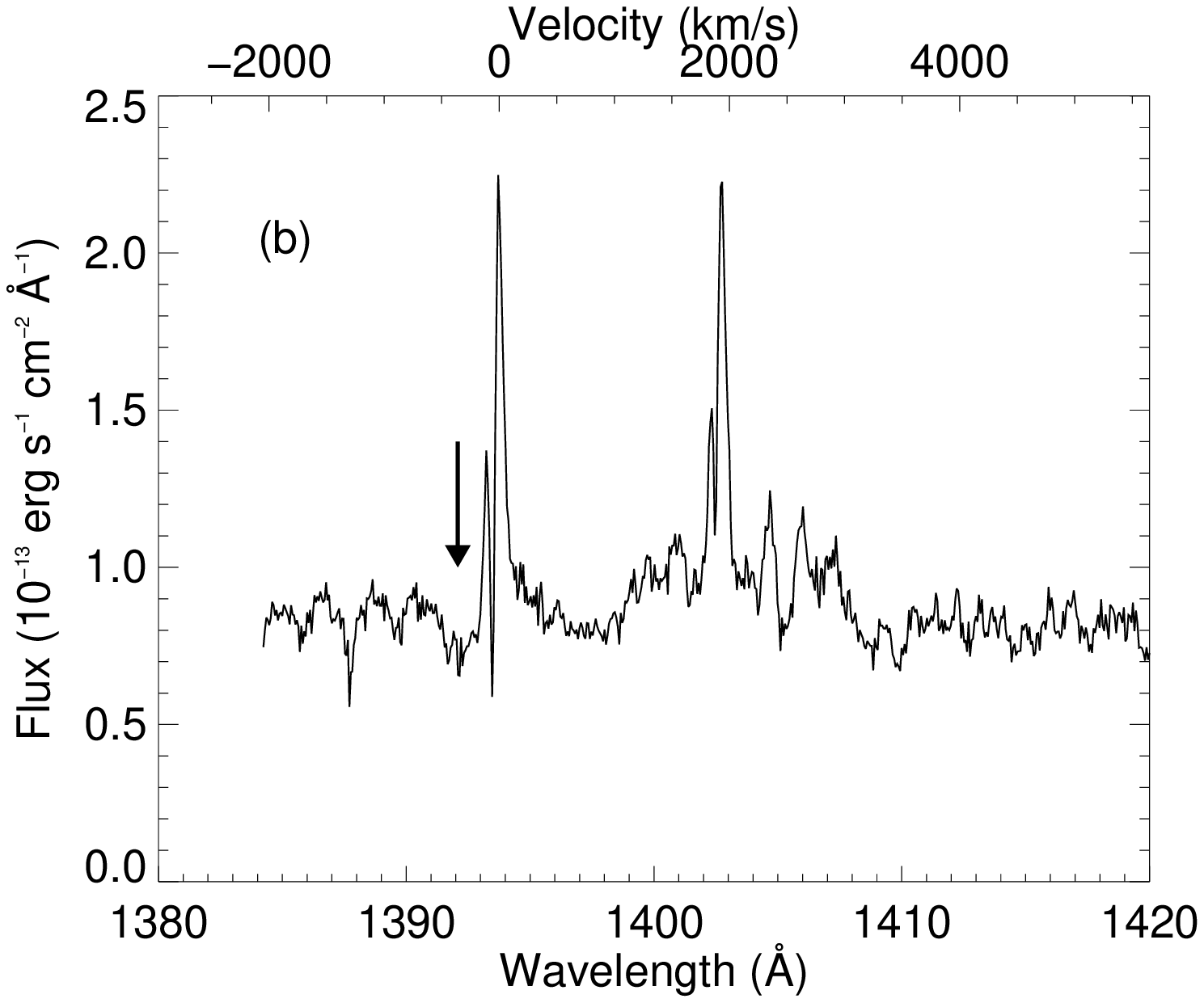}\\
\end{figure}

\setcounter{figure}{3}

\begin{figure}
\caption{}
\plotone{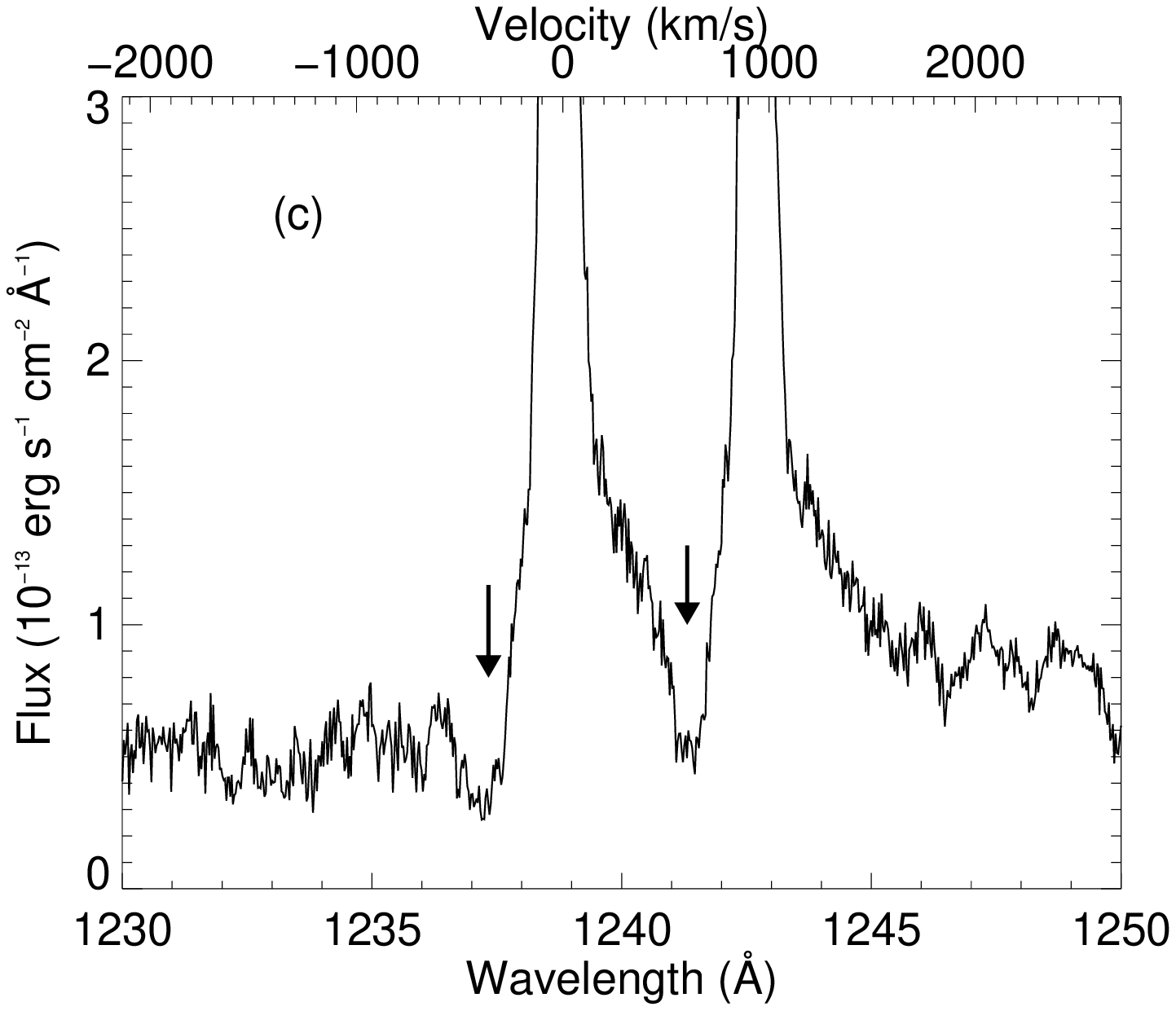}
\end{figure}

\begin{figure}
\caption{}\plotone{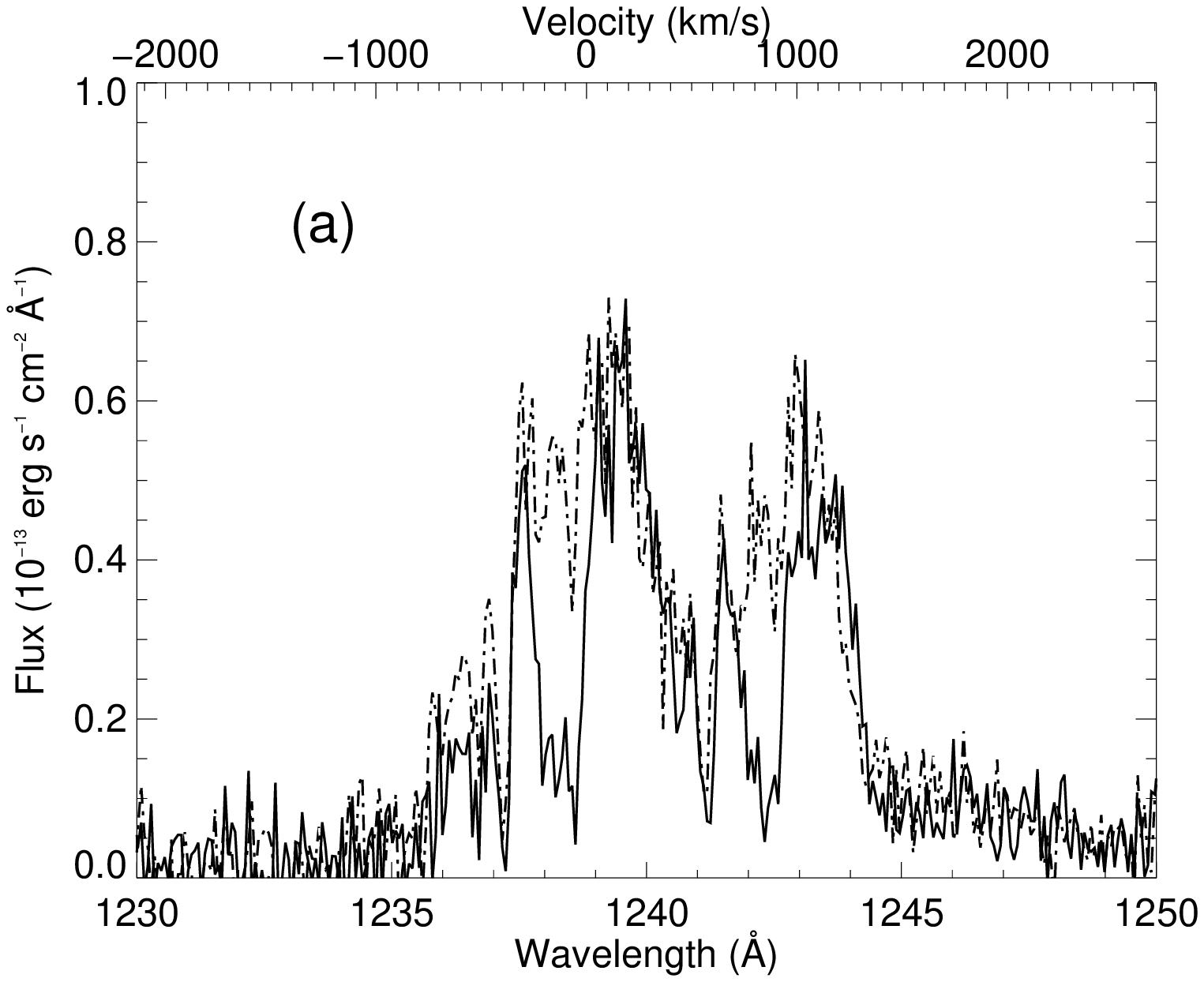}\\
\plotone{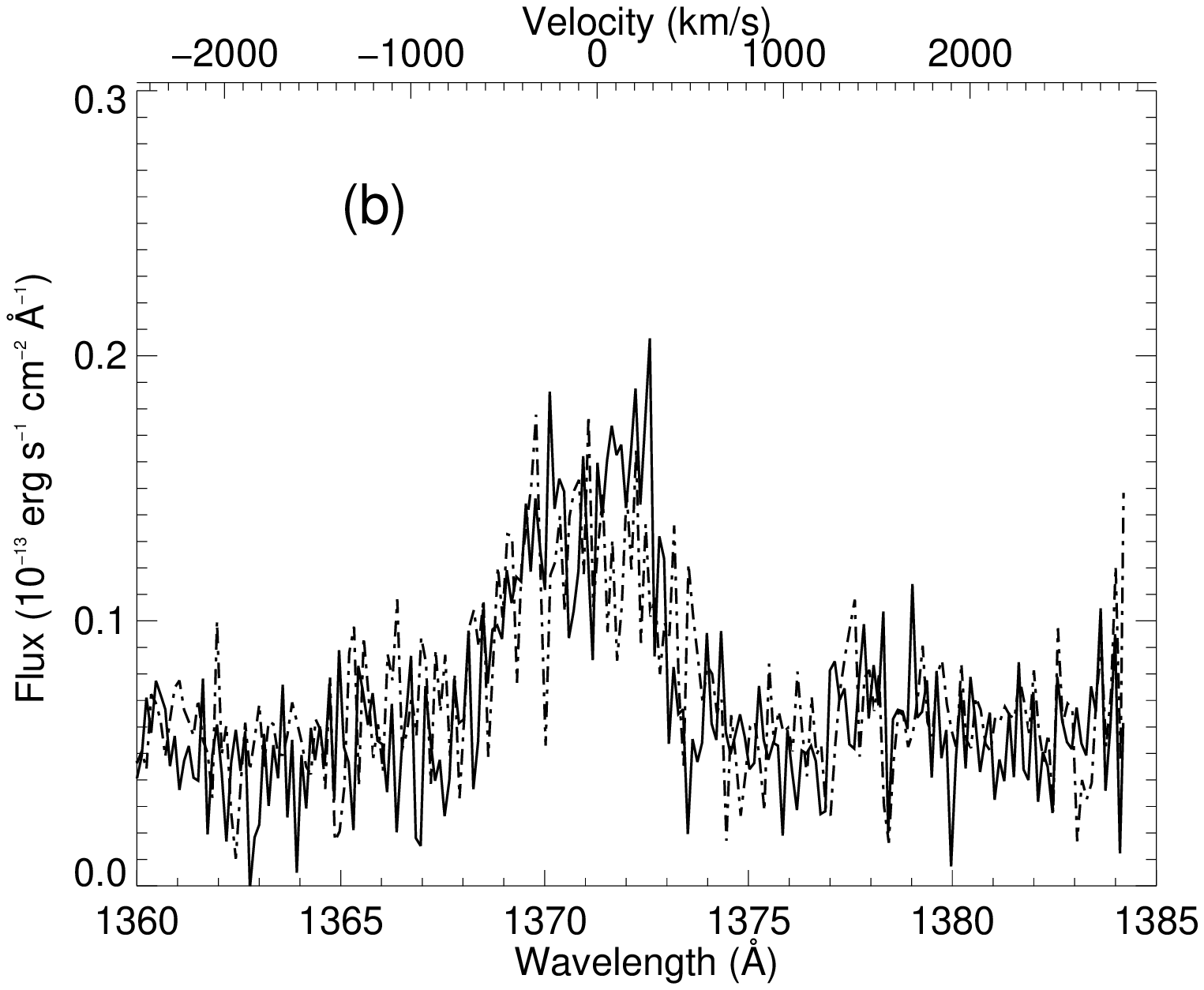}
\end{figure}

\setcounter{figure}{4}

\begin{figure}
\caption{}
\plotone{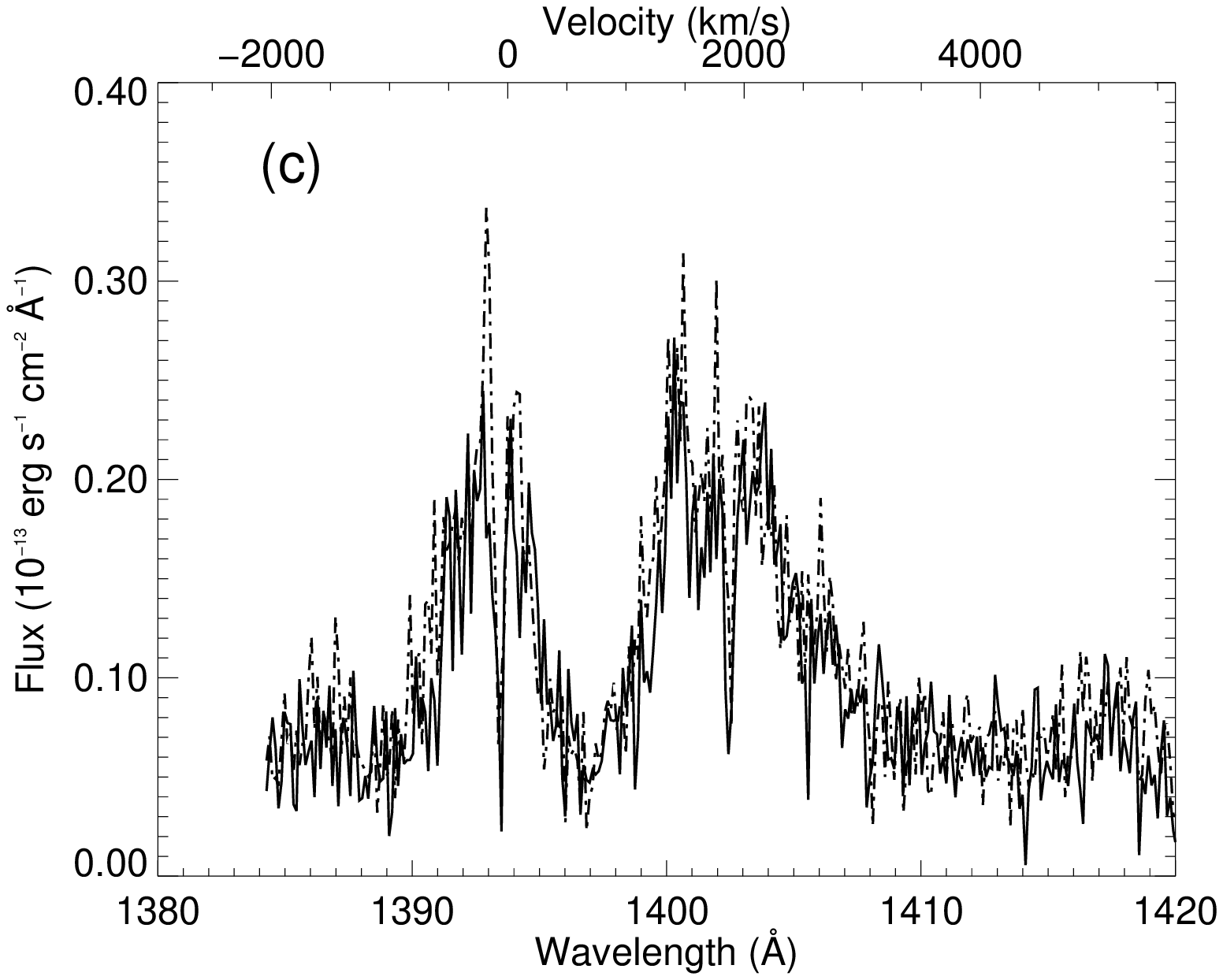}\\
\plotone{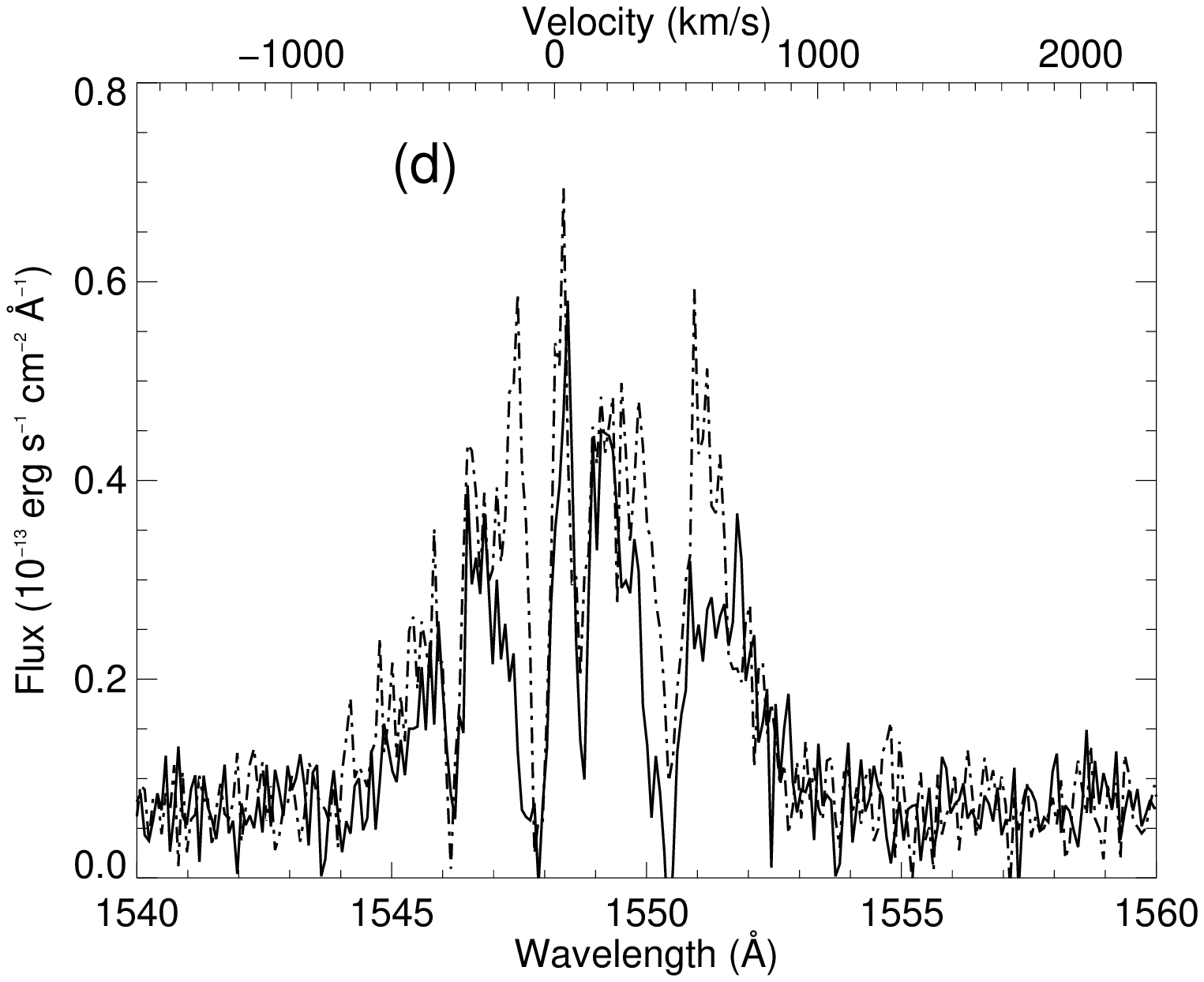}
\end{figure}

\setcounter{figure}{4}

\begin{figure}
\caption{}
\plotone{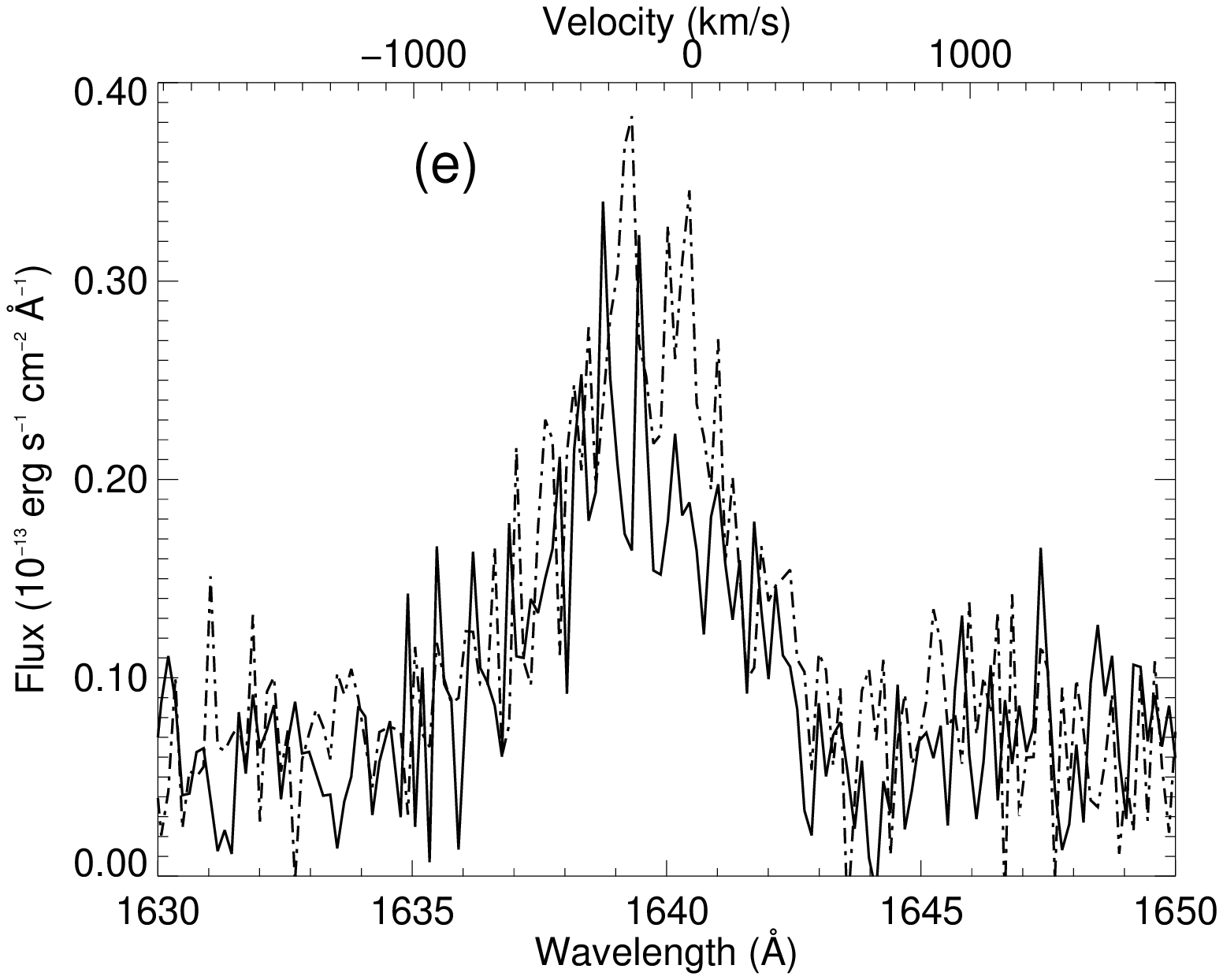}\\
\end{figure}

\begin{figure}
\caption{}
\plotone{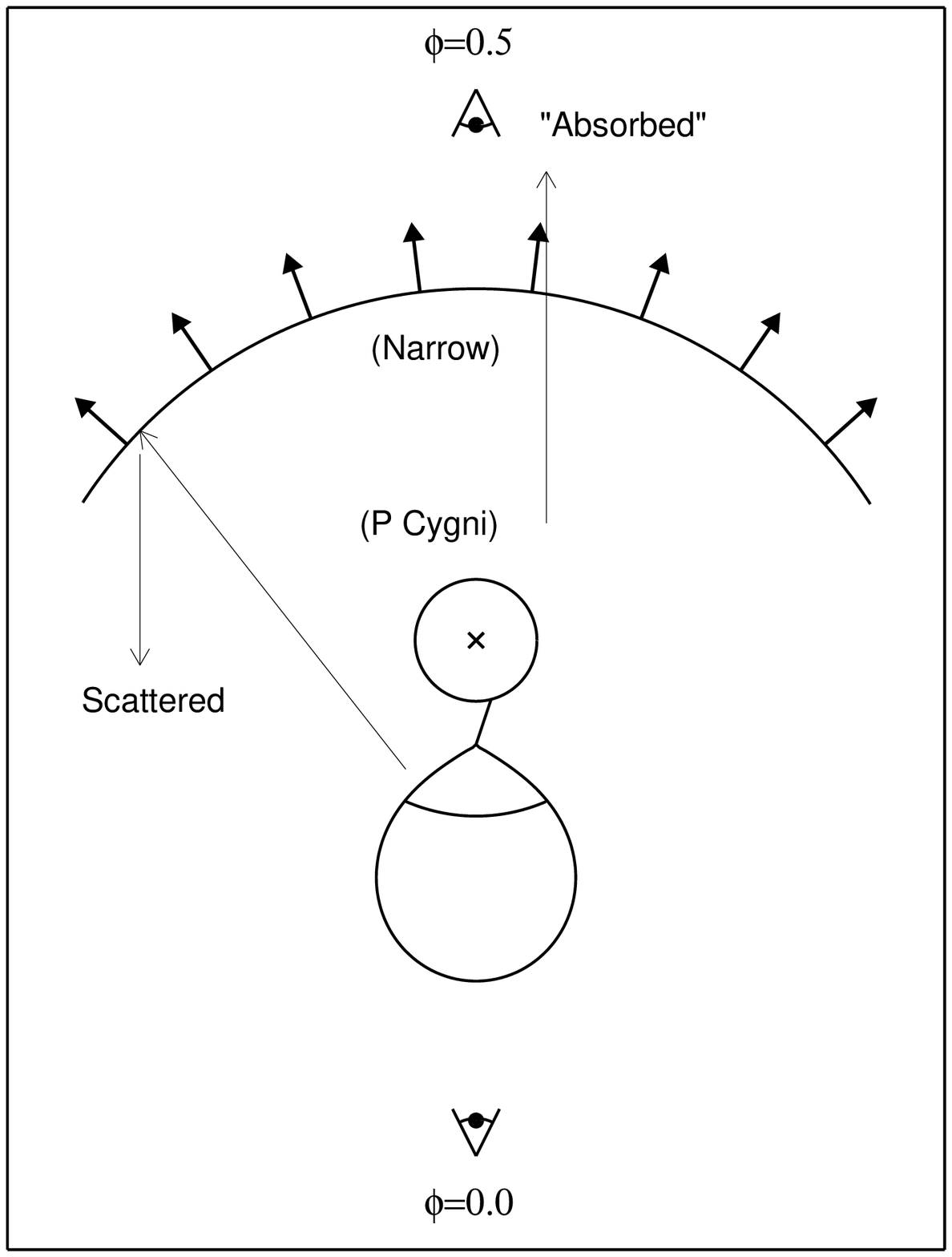}
\end{figure}

\end{document}